\documentclass[12pt]{article}

\usepackage{scicite}
\usepackage{titling}

\usepackage[colorlinks]{hyperref}

\usepackage{graphicx}
\usepackage{lscape}

\newenvironment{sciabstract}{%
\begin{quote} \bf}
{\end{quote}}

\newcounter{lastnote}

\newlength{\VSpaceBeforeTabBib}
\newlength{\VSpaceBeforeTabFoot}
\newcommand\tablefoot[1]{\VSpaceBeforeTabBib=1ex%
  \par\vspace{\VSpaceBeforeTabFoot}
  \noindent
  \begin{minipage}{\linewidth}
    {\small\bfseries Notes.}~%
    \small
    \ignorespaces
    #1%
  \end{minipage}%
}

\title{A multiple planet system of super-Earths orbiting the brightest red dwarf star GJ887}

\author
{S.~V.~Jeffers$^{1\ast}$, S.~Dreizler$^{1}$, J.~R.~Barnes$^{2}$, C.~A.~Haswell$^{2}$,   R.~P.~Nelson$^{3}$,\and E.~Rodríguez$^{4}$,   M.~J.~L\'opez-Gonz\'alez$^{4}$, N.~Morales$^{4}$, R.~Luque$^{5,6}$, \and M.~Zechmeister$^{1}$,  S.~S.~Vogt$^{7}$, J.~S.~Jenkins$^{8,9}$,   E.~Palle$^{5,6}$, Z.~M.~Berdi\~nas$^{8}$,  \and  G.~A.~L.~Coleman$^{3,10}$, M.~R.~D\'iaz$^{8}$, I.~Ribas$^{11,12}$, H.~R.~A.~Jones$^{13}$,  \and R.~P.~Butler$^{14}$,  C.~G.~Tinney$^{15}$, J.~Bailey$^{15}$,  B.~D.~Carter$^{16}$, S.~O'Toole$^{17}$, \and R.~A.~Wittenmyer$^{18}$, J.~D.~Crane$^{19}$,   F.~Feng$^{14}$,  S.~A.~Shectman$^{19}$, \and J.~Teske$^{19}$, A.~Reiners$^{1}$, P.~J.~Amado$^{4}$, G.~Anglada-Escud\'e$^{3,11,12}$\\
}

\date{}

\begin{document} 


\baselineskip24pt


\maketitle

\begin{center}
\linespread{1.15}\normalsize{{ $^{1}$ Institut f\"ur Astrophysik, Georg-August-Universit\"at, 37077 G\"ottingen, Germany \\
$^{2}$School of Physical Sciences, The Open University, Milton Keynes, MK7 6AA, UK\\
$^{3}$ School of Physics and Astronomy, Queen Mary University of London, \\E1 4NS London, UK \\
$^{4}$ Instituto de Astrof\'isica de Andaluc\'ia (Consejo Superior de Investigaciones Cient\'ificas) 18008 Granada, Spain\\
$^{5}$ Instituto de Astrofísica de Canarias, 38205 La Laguna, Tenerife, Spain\\
$^{6}$ Departamento de Astrofísica, Universidad de La Laguna, 38206 La Laguna, Tenerife, Spain\\
$^{7}$ U. of California/Lick Observatory, U. of California at Santa Cruz. Santa Cruz, CA. 95064, USA\\
$^{8}$ Departamento de Astronomia, Universidad de Chile, Santiago,Chile \\
$^{9}$ Centro de Astrof\'isica y Tecnolog\'ias Afines, Santiago, Chile \\
$^{10}$ Physikalisches Institut, Universit\"at Bern, 3012 Bern, Switzerland \\
$^{11}$ Institut de Ci\`encies de l’Espai (Consejo Superior de Investigaciones Cient\'ificas), Campus Universitat Aut\`onoma de Barcelona, E-08193 Bellaterra, Spain\\
$^{12}$ Istitut d’Estudis Espacials de Catalunya, E-08034 Barcelona, Spain\\
$^{13}$ Centre for Astrophysics Research, University of Hertfordshire, Hatfield AL10 9AB, UK \\
$^{14}$ Earth and Planets Laboratory, Carnegie Institution for Science, Washington DC 20015, USA \\
$^{15}$ Exoplanetary Science at University of New South Wales, School of Physics, University of New South Wales, Sydney 2052, Australia\\
$^{16}$ Centre for Astrophysics, University of Southern Queensland, Springfield Central QLD 4300, Australia \\
$^{17}$ Australian Astronomical Optics, Macquarie University, North Ryde NSW 2113, Australia\\
$^{18}$ Centre for Astrophysics, University of Southern Queensland, Toowoomba, QLD 4350 Australia\\
$^{19}$ The Observatories of the Carnegie Institution for Science, Pasadena, CA  91101, USA\\

}}
\normalsize{$^\ast$To whom correspondence should be addressed; E-mail: sandrajeffers.astro@gmail.com}
\end{center}


\begin{sciabstract}The nearest exoplanets to the Sun are our best possibilities for detailed characterization.  We report the discovery of a compact multi-planet system of super-Earths orbiting the nearby red dwarf GJ\,887, using radial velocity measurements.  The planets have orbital periods of 9.3 and 21.8~days. Assuming an Earth-like albedo, the equilibrium temperature of the 21.8~day planet is $\sim$350\,K; which is interior, but close to the inner edge, of the liquid-water habitable zone. We also detect a further unconfirmed signal with a period of $\sim$50\,days which could correspond to a third super-Earth in a more temperate orbit. GJ~887 is an unusually magnetically quiet red dwarf with a photometric variability below 500 parts-per-million, making its planets amenable to phase-resolved photometric characterization.
\end{sciabstract}

\section*{Main text}

At visible wavelengths, GJ~887 is the brightest red dwarf in the sky (www.recons.org) and at a distance of 3.29 parsecs (pc), the 12th closest star system to the Sun.  GJ~887 is the most massive red dwarf within 6 pc of the Sun, close enough for a direct stellar radius measurement using interferometry\cite{Boyajian2012ApJ...757..112B}.   GJ~887's stellar parameters are listed in Table~\ref{t-stparam}.   Red dwarfs are amenable to radial velocity (RV) searches for temperate Earth-mass exoplanets: their low luminosity means temperate planets have short orbital periods, and their low stellar mass implies Earth-mass planets can impart a reflex RV detectable with current instrumentation.  While the transit method of planet discovery efficiently detects planets because many stars can be simultaneously monitored, it will detect only planets that pass through the line of sight between us and the host star.  Consequently,  only 1-2\% of habitable zone planets, i.e. those with surfaces that can support liquid water, are detectable with the transit method.  The RV method is the only way to achieve a complete census of the planets orbiting our closest stellar neighbours, especially around red dwarfs. \\

We monitored GJ~887 as part of the Red Dots \#2 project. Nightly observations were taken with the High Accuracy Radial velocity Planet Searcher (HARPS) \cite{Mayor2003Msngr.114...20M} for three months.  We also obtained contemporaneous photometric observations \cite{supplmaterials}. Regular nightly sampling combined with photometric observations mitigates against false-positive exoplanet detections from intrinsic stellar variability and other sources of correlated noise.  We supplement our data with over 200 archival observations with HARPS, the Planet Finder Spectrograph (PFS) \cite{Crane2010SPIE.7735E..53C}, the High Resolution Echelle Spectrometer (HIRES) \cite{Vogt1994SPIE.2198..362V}, and the University College London Echelle Spectrograph (UCLES) \cite{Diego1990SPIE.1235..562D}, spanning nearly 20 years \cite{supplmaterials}. We used photometry from various ground-based observatories and the Transiting Exoplanet Survey Satellite mission (TESS) spacecraft \cite{Ricker2015JATIS...1a4003R}.  Tables ~S\ref{tab:data_properties} and ~S\ref{tab:phot-obslog} list all data used.  \\

\begin{figure*}[p]
    \centering
    \includegraphics[width=\textwidth]{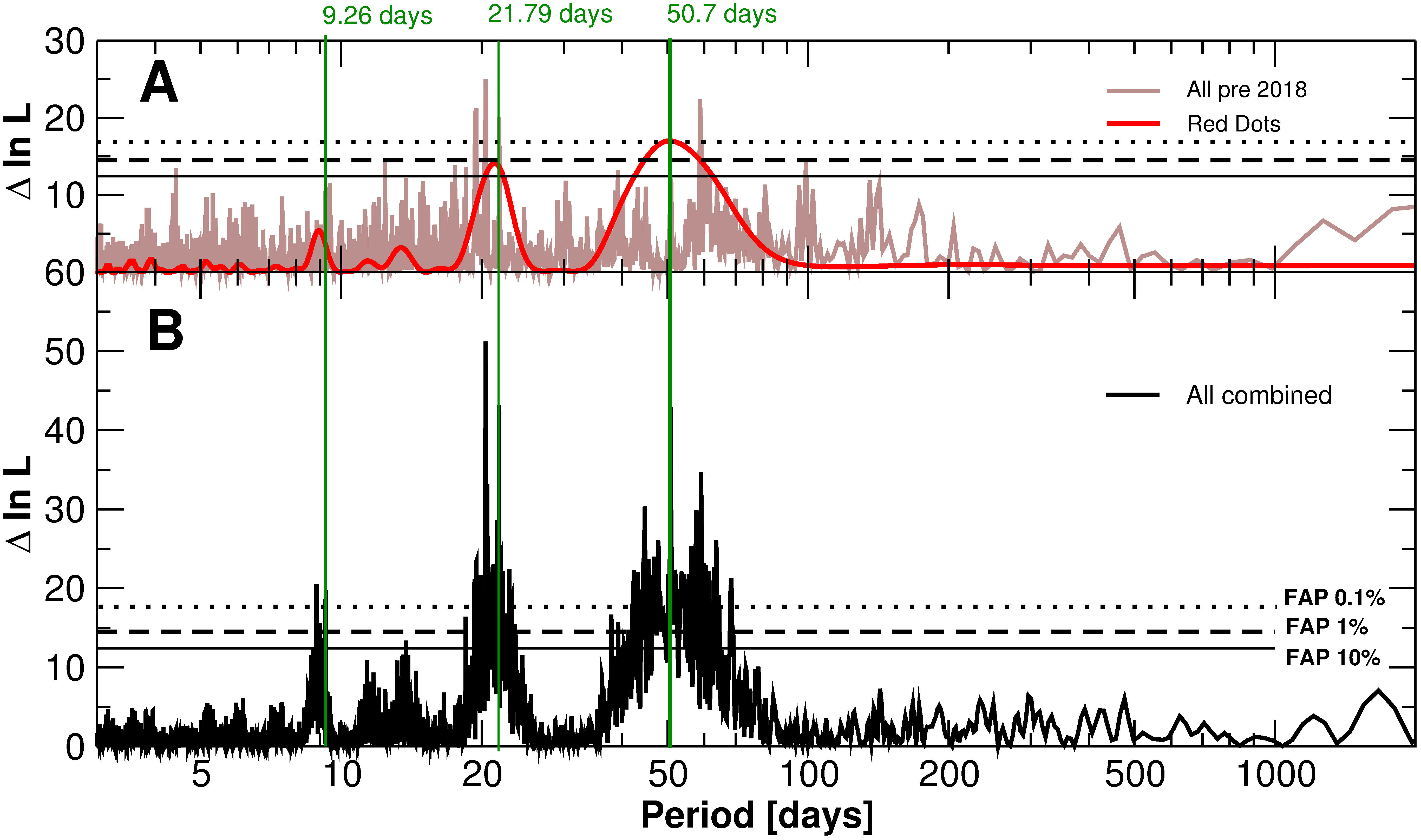}
    \caption{\label{fig:detection} {\bf{Periodograms of RV data.}} (A) is the log-likelihood periodogram ($\Delta$ In L) as obtained for all RV data before 2018 (brown) and the Red Dots $\#$2 campaign (red) analyzed separately.  (B) shows the same search for a first signal when combining all the RV observations together.  The vertical green lines indicate our derived model periods for planets b and c, and the third signal or candidate planet d.  The horizontal dashed lines in both panels indicate the False Alarm Probability (FAP) values. }
\end{figure*}

We searched for a candidate planet by adding a (circular) Keplerian orbit test signal to our base model and measuring the improvement in the logarithm of the likelihood statistic.   Our base model is composed of an offset and an instrumental jitter added to the measurement uncertainties for each data-set.
We use this to generate log-likelihood periodograms for both the RV and photometric data then search for signals by plotting the increase in the log-likelihood statistic against test period (see Fig.~1).  The highest peaks were evaluated for statistical significance \cite{Baluev2009MNRAS.393..969B, Ribas2018Natur.563..365R}. We recursively added further planet test signals, adjusting all the parameters to  maximise the likelihood for all planet signals and the parameters of the base model. We continue this iterative process until no signals below a threshold of 0.1\%~false-alarm probability are found in the time-series. We detected periodic signals at 9.3, 21.8, and 50.7~days, as shown in Figure~\ref{fig:detection}, and verified them using several independent fitting procedures and algorithm implementations \cite{supplmaterials}.    Also shown in Figure~\ref{fig:detection} is how the regular sampling of the RedDots \# 2 data set helps  disentangle the signals under investigation. \\

Stellar magnetic activity can induce an asymmetric distortion of the spectral lines, shifting the measured line centre and consequently inducing an apparent RV shift, which may appear as a false-positive exoplanet at the stellar rotation period \cite{Jeffers2014MNRAS.438.2717J}.
The rotation period of GJ~887 is unknown so we searched for periodicities in  the photometric data \cite{supplmaterials}. The archival data from 2002 - 2004 show a $\sim$200\,d period, but this was undetectable in the 2018 quasi-simultaneous photometric observations as the time span is too short.  Our analysis of the photometry from the TESS mission shows very low intrinsic variability with a semi-amplitude of 240\,ppm.  It is unclear whether this is caused by systematics known to affect the TESS observations, but we use this value  as an upper limit to the intrinsic variability of  GJ~887.  The TESS variability can be explained by one starspot, or a group of starspots, with a total diameter of 0.3\% of the stellar surface, indicating that GJ~887 is slowly rotating with very few surface brightness inhomogeneities  \cite{Barnes2015ApJ...812...42B}.  The combination of this very low spot coverage and photometric variability, its value of
log($R^{'}_{HK}$), a metric derived from stellar Ca II H\& K lines, of -4.805 \cite{BoroSaikia2018A&A...616A.108B}, and that GJ~887 has a very low H$\alpha$ activity \cite{Jeffers2018A&A...614A..76J}, makes it less magnetically active than most stars with the same effective temperature.\\ 

Given that the detected RV signals are clear in the Red Dots \# 2 HARPS spectra alone, we investigated additional spectral signatures of stellar magnetic activity of this data set. We extracted a time series of the flux in the cores of the NaD, H$\alpha$ and H$\beta$ lines;  and the S-index, this being the ratio of flux in the cores in the Ca II H\& K lines compared to the continuum (see \cite{supplmaterials} for further details). The S-index and Na\,D lines both show a weak signal at about 55\,d, while the H$\alpha$ and H$\beta$ lines show a weak signal at 38 days. These differing periodicities could reflect timescales of various stellar activity processes on the star, and despite being low in amplitude, they make a planetary origin for RV signals in the 30-60 days domain less certain. None of these periodicies in activity are close to the RV signals at 9.3\,d and 21.8\,d day but question the RV signal detected at 50.7\,d.\\


Correlated noise, e.g. caused by stellar activity, can be assessed via the covariances between observations. To further verify the planetary origin of the detected RV signals we fitted maximum likelihood model functions using two planet models with and without Gaussian Processes \cite[GP]{supplmaterials}. All of the models including GP improved the fit to the data compared to those without, and the amplitude of the signals with periods of 9.3\,d and 21.8\,d remained unchanged within their $1\sigma$ uncertainty. The modelling of the correlated noise using GP therefore does not affect these two signals. However,the significance of the third signal drops significantly when including a GP in the model, casting further doubts on its Keplerian nature. Table \ref{tab:modelcomparison} in supplementary materials shows the derived values and relevant statistical quantities of the preferred final model.\\

\begin{figure*}[p]
    \centering
    \includegraphics[width=\textwidth]{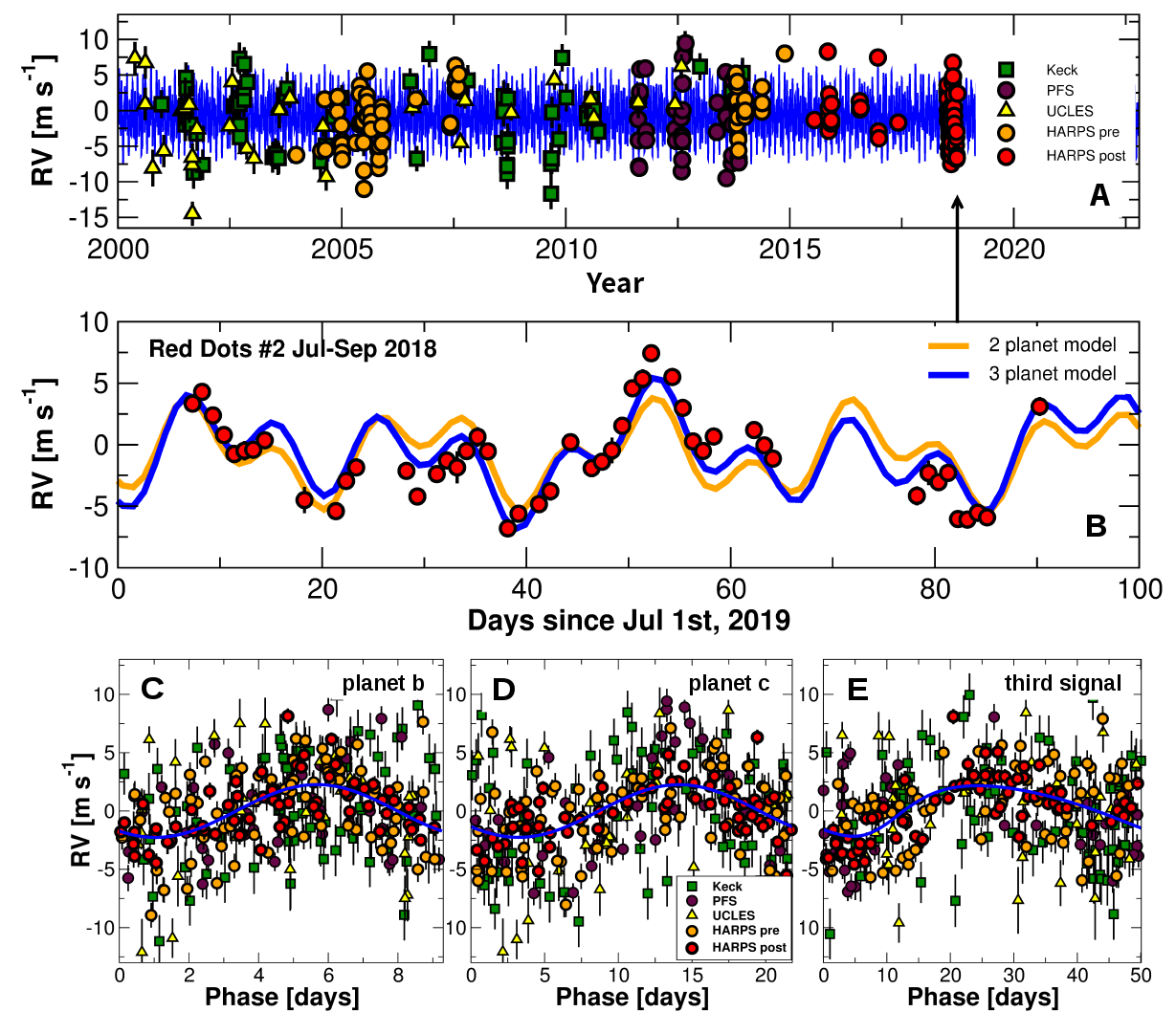}
    \caption{\label{fig:rvs} {\bf{Time series of radial velocity measurements.}} All radial velocity measurements with instruments used indicated in the key where HARPS-pre and HARPS-post refer to data collected before and after the fibre upgrade. (A): Radial velocity measurements of GJ~887 over 18 years using different instruments as indicated.  The best fit model with three Keplerian signals is shows as a solid blue line. (B): Zoom in on panel showing the Red Dots \#2 observations.  The vertical green lines indicate our derived model periods for planets b and c, and the candidate planet d.  A planetary origin for the $\sim$50 day signal is uncertain, but three periodic modulations are required to fit the observations. Panels C, D to E:  Data are folded on the period of each candidate signal after subtracting the other signals. Each panel shows the best fit model signal as a blue solid line.}
\end{figure*}

We conclude that the two signals with orbital periods of 9.3~days and 21.8~days correspond to two exoplanets, planet~b and planet~c. The minimum masses are 4.2$\pm 0.6 $\,Earth masses (${\rm M}_{\oplus}$), 7.6\,$\pm 1.2\, {\rm M}_{\oplus}$, i.e. two super-Earth exoplanets which orbit at semi-major axes of 0.068 astronomical units (au) and 0.120\,au.  The inner planet has an orbital eccentricity consistent with zero as shown in Fig.~\ref{fig:cornerPlanetb}, but the outer planet is more likely to have low but non-zero eccentricity  (Fig.~\ref{fig:cornerPlanetc}). We regard the third signal at approximately $\sim$50 days (c.f. Fig.~\ref{fig:rvs}) as dubious and likely related to stellar activity. The fits to our two-planet model, and the two-planets + third signal model are shown in Figure~\ref{fig:rvs}.  \\

The long term dynamical stability of the orbits can also be used to further test the physical reality of a system, and investigate the possible presence of dynamically interesting configurations such as dynamical resonances. We perform this dynamical stability study using {\sc mercury6 \cite{Chambers1999MNRAS.304..793C}}. 
We find that all two-planet solutions are stable even if eccentricities are left unconstrained. The ratio of periods of these to planets is close to 7:3, but the simulations do not support the existence of a dynamical resonance based on the absence of oscillating orbital alignment variations \cite{perryman2018exha.book.....P}. We find, however, that the system must be in a dynamically active state driving oscillatory changes in the eccentricities of both planets. These interactions produce very regular variations which support the hypothesis that the two planet configuration is dynamically stable on very long time-scales.  Concerning a putative system with three planets, only about 25\% of our best one thousand fits would be dynamically stable over 10$^5$\,yr, but this is mostly causing by the poorly constrained eccentricities. Given that the eccentricities are only really upper limits, we checked what happens when orbits are assumed circular (initial zero eccentricities). Even in this three-planet case, more than 99\% of configurations were found to be stable, meaning that the presence of a third planet cannot be ruled out using dynamic stability considerations.\\

The separations between the planets, in units of their spheres of gravitational influence or Hill radii, are $\sim 19.1$ for planets b and c, and $\sim 17.2$ for planets c and d (assuming planet d is real and has a mass of 8.3 M$_{\oplus}$); these values are consistent with the system having undergone dynamical relaxation \cite{2015ApJ...807...44P}. Dynamical relaxation in systems of super-Earths results in $\sim 80\%$ of planets having orbital eccentricities $e_{\rm p} \le 0.1$, with the remaining 20\% having $e_{\rm p} \le 0.3$ \cite{2020MNRAS.491.5595P}. We examined the tidal evolution of GJ~887-b using analytical methods \cite{2004ApJ...610..464D,1998ApJ...499..853E} finding that the tidal circularization time scale of GJ~887-b
is a few Gyr for an assumed tidal dissipation parameter $Q^{\prime}_p=1000$. This is consistent with our observation that GJ~887-b's orbit is almost circular.\\

The multi-planet super-Earth system around GJ~887 is consistent with recent planet formation models \cite{Coleman2016MNRAS.457.2480C,Lambrechts2019A&A...627A..83L}. These models typically form chains of multiple planets trapped in mean-motion resonances that then migrate into orbits close to the central star. Depending on where the initial planets formed in the protoplanetary disc, they could have accreted significant amounts of water ice or purely dry rocky silicates. As such the planets may be either water-rich or water-poor. At the end of the gas disc lifetime, the resonant chains of planets can remain stable yielding systems similar to the seven-planet TRAPPIST-1 planetary system \cite{Gillon2017Natur.542..456G} or they can become unstable, leading to collisions between planets, and thus a non-resonant configuration \cite{Coleman2016MNRAS.457.2480C}. The GJ~887 planetary system appears more consistent with the latter, unstable evolution. The existence of dynamical resonances can be very sensitive to the existence or absence of additional planets. Consequently, if the third signal at 50.7\,days is real or if there are additional planets, this may result in a more resonant system.\\ 

\begin{figure}[p]
    \centering
    \includegraphics[angle=270,width=0.9\textwidth]{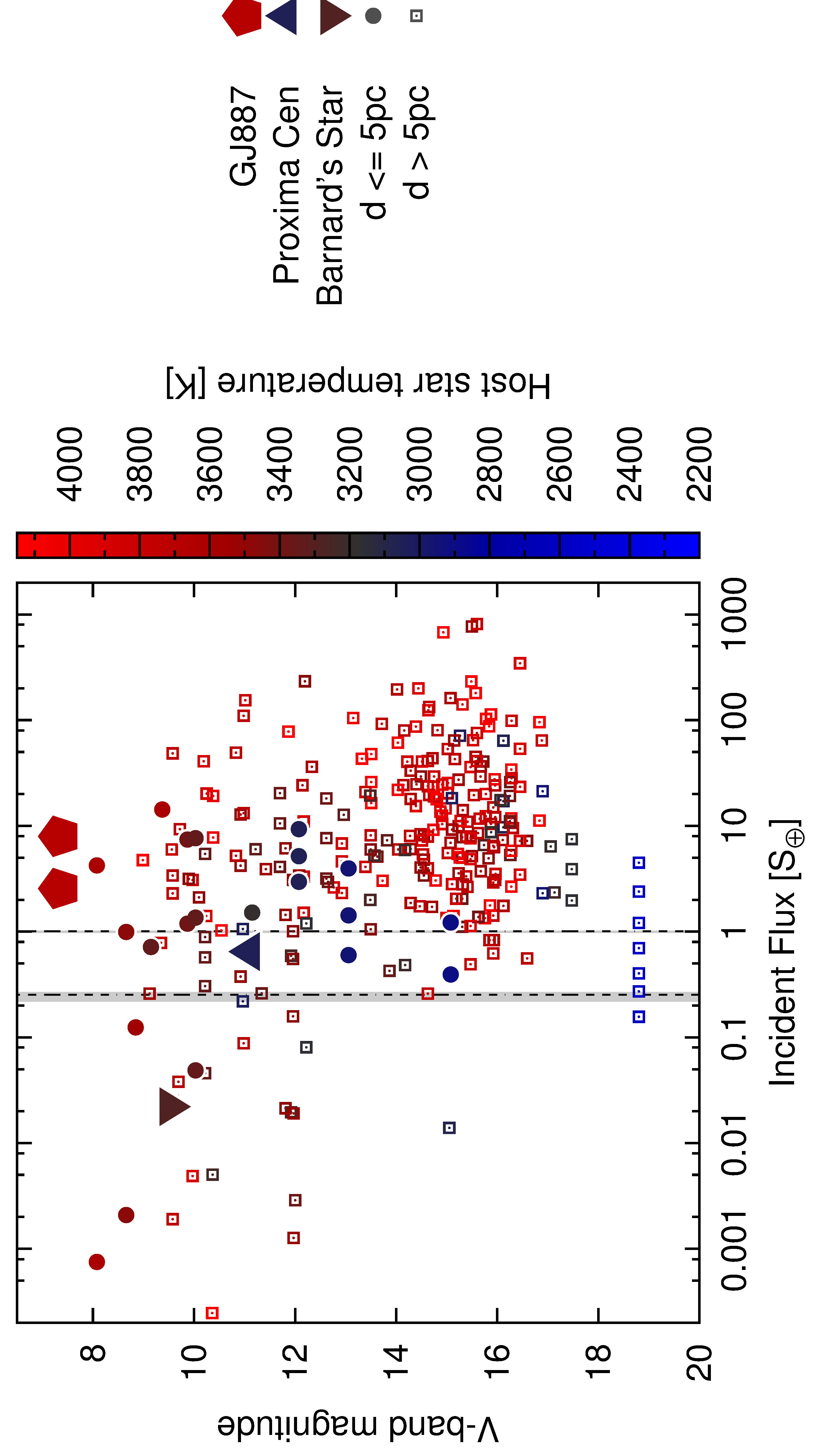}
    \caption{{\bf{The incident flux (or insolation) of planets orbiting M dwarfs.}} The dashed lines delimit the
    habitable zone around GJ\,887 for the maximum greenhouse planetary atmosphere (left) and the runaway greenhouse planetary atmosphere (right)  \cite{Kopparapu2014ApJ...787L..29K}.  The solid vertical grey lines indicate the range of limits for the host stars of all planets plotted; these stars have $T_{\rm eff}$ ranging from 2400 - 4150 K (see colour bar).  GJ~887\,b and GJ~887\,c are indicated by the large red pentagons. }
    \label{fig:P-M-GAIA}
\end{figure}

According to the calculations of \cite{Kopparapu2014ApJ...787L..29K}, the orbits of GJ~887-b and GJ~887-c could support liquid water (commonly refereed to as the star's Habitable Zone, or HZ) on their surfaces extends from approximately 0.19 au to 0.38 au. 
With a semi-major axis ($a_{\rm p}) = 0.120 \pm 0.004$, GJ~887 c is closer to its host star than the HZ, but near the inner edge.  If the $\sim50$\,d signal is planetary in origin, it corresponds to a super-Earth in GJ\,887's liquid-water HZ. Assuming an albedo, $\alpha$, similar to Earth’s ($\alpha = 0.3$), the equilibrium temperature, $T_{\rm eq}$, of the planets b and c would be 468\,K and 352\,K  respectively. Their incident energy fluxes from the star (or insolation $S$), are 7.95 and 2.56 times the Sun's insolation on the Earth. Fig.~\ref{fig:P-M-GAIA} shows the insolation of known planets orbiting M dwarfs as a function of host star apparent magnitude. GJ~887 is has the brightest apparent magnitude among all other known M dwarf planet hosts.  This combined with the high photometric stability of GJ~887, exhibited in the TESS light curves, and the high planet-star brightness and radius ratios, make these planets suitable targets for phased resolved photometric studies, especially in emission light  \cite{Kempton2018PASP..130k4401K}.  Similarly, spectrally resolved phase photometry has been shown to be able to uncover the presence of an atmosphere and of molecules such as CO$_2$ (e.g.~\cite{Snellen2017AJ....154...77S}). \\ 

\bibliography{scibib}

\begin{thebibliography}{10}

\bibitem{Boyajian2012ApJ...757..112B}
T.~S. {Boyajian}, {\it et~al.\/}, {\it \apj\/} {\bf 757}, 112 (2012).

\bibitem{Mayor2003Msngr.114...20M}
M.~{Mayor}, {\it et~al.\/}, {\it The Messenger\/} {\bf 114}, 20 (2003).

\bibitem{supplmaterials}
{See supplementary materials.}

\bibitem{Crane2010SPIE.7735E..53C}
J.~D. {Crane}, {\it et~al.\/}, {\it Ground-based and Airborne Instrumentation
  for Astronomy III\/} (2010), vol. 7735 of {\it \procspie\/}, p. 773553.

\bibitem{Vogt1994SPIE.2198..362V}
S.~S. {Vogt}, {\it et~al.\/}, {\it Instrumentation in Astronomy VIII\/}, D.~L.
  {Crawford}, E.~R. {Craine}, eds. (1994), vol. 2198 of {\it \procspie\/}, p.
  362.

\bibitem{Diego1990SPIE.1235..562D}
F.~{Diego}, A.~{Charalambous}, A.~C. {Fish}, D.~D. {Walker}, {\it
  Instrumentation in Astronomy VII\/}, D.~L. {Crawford}, ed. (1990), vol. 1235
  of {\it \procspie\/}, pp. 562--576.

\bibitem{Ricker2015JATIS...1a4003R}
G.~R. {Ricker}, {\it et~al.\/}, {\it Journal of Astronomical Telescopes,
  Instruments, and Systems\/} {\bf 1}, 014003 (2015).

\bibitem{Baluev2009MNRAS.393..969B}
R.~V. {Baluev}, {\it \mnras\/} {\bf 393}, 969 (2009).

\bibitem{Ribas2018Natur.563..365R}
I.~{Ribas}, {\it et~al.\/}, {\it \nat\/} {\bf 563}, 365 (2018).

\bibitem{Jeffers2014MNRAS.438.2717J}
S.~V. {Jeffers}, {\it et~al.\/}, {\it \mnras\/} {\bf 438}, 2717 (2014).

\bibitem{Barnes2015ApJ...812...42B}
J.~R. {Barnes}, {\it et~al.\/}, {\it \apj\/} {\bf 812}, 42 (2015).

\bibitem{BoroSaikia2018A&A...616A.108B}
S.~{Boro Saikia}, {\it et~al.\/}, {\it \aap\/} {\bf 616}, A108 (2018).

\bibitem{Jeffers2018A&A...614A..76J}
S.~V. {Jeffers}, {\it et~al.\/}, {\it \aap\/} {\bf 614}, A76 (2018).

\bibitem{Chambers1999MNRAS.304..793C}
J.~E. {Chambers}, {\it \mnras\/} {\bf 304}, 793 (1999).

\bibitem{perryman2018exha.book.....P}
M.~{Perryman}, {\it {The Exoplanet Handbook}\/} (Cambridge University Press,
  2018).

\bibitem{2015ApJ...807...44P}
B.~{Pu}, Y.~{Wu}, {\it \apj\/} {\bf 807}, 44 (2015).

\bibitem{2020MNRAS.491.5595P}
S.~T.~S. {Poon}, R.~P. {Nelson}, S.~A. {Jacobson}, A.~{Morbidelli}, {\it
  \mnras\/} {\bf 491}, 5595 (2020).

\bibitem{2004ApJ...610..464D}
I.~{Dobbs-Dixon}, D.~N.~C. {Lin}, R.~A. {Mardling}, {\it \apj\/} {\bf 610}, 464
  (2004).

\bibitem{1998ApJ...499..853E}
P.~P. {Eggleton}, L.~G. {Kiseleva}, P.~{Hut}, {\it \apj\/} {\bf 499}, 853
  (1998).

\bibitem{Coleman2016MNRAS.457.2480C}
G.~A.~L. {Coleman}, R.~P. {Nelson}, {\it \mnras\/} {\bf 457}, 2480 (2016).

\bibitem{Lambrechts2019A&A...627A..83L}
M.~{Lambrechts}, {\it et~al.\/}, {\it \aap\/} {\bf 627}, A83 (2019).

\bibitem{Gillon2017Natur.542..456G}
M.~{Gillon}, {\it et~al.\/}, {\it \nat\/} {\bf 542}, 456 (2017).

\bibitem{Kopparapu2014ApJ...787L..29K}
R.~K. {Kopparapu}, {\it et~al.\/}, {\it \apjl\/} {\bf 787}, L29 (2014).

\bibitem{Kempton2018PASP..130k4401K}
E.~M.~R. {Kempton}, {\it et~al.\/}, {\it \pasp\/} {\bf 130}, 114401 (2018).

\bibitem{Snellen2017AJ....154...77S}
I.~A.~G. {Snellen}, {\it et~al.\/}, {\it \aj\/} {\bf 154}, 77 (2017).

\bibitem{Schweitzer2019A&A...625A..68S}
A.~{Schweitzer}, {\it et~al.\/}, {\it \aap\/} {\bf 625}, A68 (2019).

\bibitem{Mann2015ApJ...804...64M}
A.~W. {Mann}, G.~A. {Feiden}, E.~{Gaidos}, T.~{Boyajian}, K.~{von Braun}, {\it
  \apj\/} {\bf 804}, 64 (2015).

\bibitem{GAIA2018A&A...616A...1G}
{Gaia Collaboration}, {\it et~al.\/}, {\it \aap\/} {\bf 616}, A1 (2018).

\bibitem{Segransan2003A&A...397L...5S}
D.~{S{\'e}gransan}, P.~{Kervella}, T.~{Forveille}, D.~{Queloz}, {\it \aap\/}
  {\bf 397}, L5 (2003).

\bibitem{Demory2009A&A...505..205D}
B.-O. {Demory}, {\it et~al.\/}, {\it \aap\/} {\bf 505}, 205 (2009).

\bibitem{Browning2010AJ....139..504B}
M.~K. {Browning}, G.~{Basri}, G.~W. {Marcy}, A.~A. {West}, J.~{Zhang}, {\it
  \aj\/} {\bf 139}, 504 (2010).

\bibitem{Rabus2019MNRAS.484.2674R}
M.~{Rabus}, {\it et~al.\/}, {\it \mnras\/} {\bf 484}, 2674 (2019).

\bibitem{Lovis2007A&A...468.1115L}
C.~{Lovis}, F.~{Pepe}, {\it \aap\/} {\bf 468}, 1115 (2007).

\bibitem{Anglada2012ApJS..200...15A}
G.~{Anglada-Escud{\'e}}, R.~P. {Butler}, {\it \apjs\/} {\bf 200}, 15 (2012).

\bibitem{locurto2015Msngr.162....9L}
G.~{Lo Curto}, {\it et~al.\/}, {\it The Messenger\/} {\bf 162}, 9 (2015).

\bibitem{Tal-Or2019MNRAS.484L...8T}
L.~{Tal-Or}, T.~{Trifonov}, S.~{Zucker}, T.~{Mazeh}, M.~{Zechmeister}, {\it
  \mnras\/} {\bf 484}, L8 (2019).

\bibitem{Sicardy2011Natur.478..493S}
B.~{Sicardy}, {\it et~al.\/}, {\it \nat\/} {\bf 478}, 493 (2011).

\bibitem{Pojmanski1997AcA....47..467P}
G.~{Pojmanski}, {\it \actaa\/} {\bf 47}, 467 (1997).

\bibitem{Duncan1991ApJS...76..383D}
D.~K. {Duncan}, {\it et~al.\/}, {\it \apjs\/} {\bf 76}, 383 (1991).

\bibitem{Barnes2016MNRAS.462.1012B}
J.~R. {Barnes}, C.~A. {Haswell}, D.~{Staab}, G.~{Anglada-Escud{\'e}}, {\it
  \mnras\/} {\bf 462}, 1012 (2016).

\bibitem{Lucy2005A&A...439..663L}
L.~B. {Lucy}, {\it \aap\/} {\bf 439}, 663 (2005).

\bibitem{2020SciPy-NMeth}
P.~{Virtanen}, {\it et~al.\/}, {\it Nature Methods\/} {\bf 17}, 261 (2020).

\bibitem{Foreman-Mackey2013PASP..125..306F}
D.~{Foreman-Mackey}, D.~W. {Hogg}, D.~{Lang}, J.~{Goodman}, {\it \pasp\/} {\bf
  125}, 306 (2013).

\bibitem{2017AJ....154..220F}
D.~{Foreman-Mackey}, E.~{Agol}, S.~{Ambikasaran}, R.~{Angus}, {\it \aj\/} {\bf
  154}, 220 (2017).

\bibitem{Press1996}
W.~H. Press, S.~a. Teukolsky, W.~T. Vetterling, B.~P. Flannery, {\it {Numerical
  Recipes in Fortran 77: the Art of Scientific Computing. Second Edition}\/},
  vol.~1 (1996).

\bibitem{McQuillan2014ApJS..211...24M}
A.~{McQuillan}, T.~{Mazeh}, S.~{Aigrain}, {\it \apjs\/} {\bf 211}, 24 (2014).

\bibitem{Newton2018AJ....156..217N}
E.~R. {Newton}, N.~{Mondrik}, J.~{Irwin}, J.~G. {Winters}, D.~{Charbonneau},
  {\it \aj\/} {\bf 156}, 217 (2018).

\bibitem{AstudilloDefru2017A&A...600A..13A}
N.~{Astudillo-Defru}, {\it et~al.\/}, {\it \aap\/} {\bf 600}, A13 (2017).

\bibitem{Jenkins2019MNRAS.487..268J}
J.~S. {Jenkins}, {\it et~al.\/}, {\it \mnras\/} {\bf 487}, 268 (2019).

\bibitem{2009PASP..121.1016M}
S.~{Meschiari}, {\it et~al.\/}, {\it \pasp\/} {\bf 121}, 1016 (2009).

\bibitem{juliet2019MNRAS.tmp.2366E}
N.~{Espinoza}, D.~{Kossakowski}, R.~{Brahm}, {\it \mnras\/} p. 2366 (2019).

\bibitem{Trifonov2019ascl.soft06004T}
T.~{Trifonov}, {\it Astrophysics Source Code Library\/} p. ascl:1906.004
  (2019).

\bibitem{Zechmeister2009A&A...496..577Z}
M.~{Zechmeister}, M.~{K{\"u}rster}, {\it \aap\/} {\bf 496}, 577 (2009).

\bibitem{Luque2019A&A...628A..39L}
R.~{Luque}, {\it et~al.\/}, {\it \aap\/} {\bf 628}, A39 (2019).

\bibitem{Wu2019ApJ...874...91W}
Y.~{Wu}, {\it \apj\/} {\bf 874}, 91 (2019).

\bibitem{2019hsax.conf..266B}
Z.~M. {Berdi{\~n}as}, {\it et~al.\/}, {\it Highlights on Spanish Astrophysics
  X\/} (2019), pp. 266--271.

\bibitem{Gladman1993Icar..106..247G}
B.~{Gladman}, {\it Icarus\/} {\bf 106}, 247 (1993).

\bibitem{2013MNRAS.436.3547G}
C.~A. {Giuppone}, M.~H.~M. {Morais}, A.~C.~M. {Correia}, {\it \mnras\/} {\bf
  436}, 3547 (2013).

\bibitem{2017A&A...605A..72L}
J.~{Laskar}, A.~C. {Petit}, {\it \aap\/} {\bf 605}, A72 (2017).

\bibitem{1996Icar..119..261C}
J.~E. {Chambers}, G.~W. {Wetherill}, A.~P. {Boss}, {\it \icarus\/} {\bf 119},
  261 (1996).

\bibitem{2002Icar..156..570M}
F.~{Marzari}, S.~J. {Weidenschilling}, {\it \icarus\/} {\bf 156}, 570 (2002).

\bibitem{2013ApJ...775...53H}
B.~M.~S. {Hansen}, N.~{Murray}, {\it \apj\/} {\bf 775}, 53 (2013).

\bibitem{1966Icar....5..375G}
P.~{Goldreich}, S.~{Soter}, {\it \icarus\/} {\bf 5}, 375 (1966).

\bibitem{1981Icar...47....1Y}
C.~F. {Yoder}, S.~J. {Peale}, {\it \icarus\/} {\bf 47}, 1 (1981).

\bibitem{2016CeMDA.126..145L}
V.~{Lainey}, {\it Celestial Mechanics and Dynamical Astronomy\/} {\bf 126}, 145
  (2016).

\bibitem{1998ApJ...502..788T}
C.~{Terquem}, J.~C.~B. {Papaloizou}, R.~P. {Nelson}, D.~N.~C. {Lin}, {\it
  \apj\/} {\bf 502}, 788 (1998).

\bibitem{Teske2016AJ....152..167T}
J.~K. {Teske}, {\it et~al.\/}, {\it \aj\/} {\bf 152}, 167 (2016).

\bibitem{Butler_2017}
R.~P. Butler, {\it et~al.\/}, {\it The Astronomical Journal\/} {\bf 153}, 208
  (2017).

\bibitem{Tinney2001ApJ...551..507T}
C.~G. {Tinney}, {\it et~al.\/}, {\it \apj\/} {\bf 551}, 507 (2001).

\bibitem{Weiss2013ApJ...768...14W}
L.~M. {Weiss}, {\it et~al.\/}, {\it \apj\/} {\bf 768}, 14 (2013).

\end{thebibliography}

\bibliographystyle{Science}

{\noindent\bf{Acknowledgements:}} 
We would like to kindly thank P.~A.~Pe\~na Rojas for contributing results using the {\sc{EMPEROR}} code. Based on observations collected at the European Organisation for Astronomical Research in the Southern Hemisphere under ESO programmes 101.C-0516, 101.C-0494 and 102.C-0525. This paper includes data gathered with the 6.5 meter Magellan Telescopes located at Las Campanas Observatory, Chile. Photometric data were partly collected with the robotic 40-cm telescope ASH2 at the SPACEOBS observatory (San Pedro de Atacama, Chile) operated by the Instituto de Astrofísica de Andalucía (IAA). This paper includes data collected with the TESS mission, obtained from the MAST data archive at the Space Telescope Science Institute (STScI). Funding for the TESS mission is provided by the NASA Explorer Program. STScI is operated by the Association of Universities for Research in Astronomy, Inc., under NASA contract NAS 5–26555. This paper includes data gathered with the 6.4 meter Magellan Telescopes located at Las Campanas Observatory, Chile.  \\

{\noindent\bf{Funding:}}
SVJ acknowledges the support of the  German Science Foundation (DFG) Research Unit FOR2544 `Blue Planets around Red Stars', project JE 701/3-1 and DFG priority program SPP 1992 `Exploring the Diversity of Extrasolar Planets' (RE 1664/18).  JRB and CAH acknowledge support from STFC Consolidated Grants ST/P000584/1 and ST/T000295/1. RPN was supported by STFC Consolidated Grant ST/P000592/1. 
ER, MJL-G, NM and PJA acknowledge support from the Spanish Agencia Estatal de Investigación through projects AYA2017-89637-R,
AYA2016-79425-C3-3-P, ESP2017-87676-C5-2-R, ESP2017-87143-R and the Centre of Excellence `Severo Ochoa' Instituto de Astrofísica de Andalucía (SEV-2017-0709). EP acknowledges support from the Spanish Agencia Estatal de Investigación PGC2018-098153-B-C31 and ESP2016-80435-C2-2-R.  ZMB acknowledges funds from CONICYT/FONDECYT POSTDOCTORADO 3180405. GALC acknowledges support from the Swiss National Science Foundation. MRD acknowledges support of CONICYT/PFCHA-Doctorado Nacional 21140646, Chile.  IR acknowledges support from the Spanish Ministry of Science and Innovation and the European Regional Development Fund through grants ESP2016-80435-C2-1-R and PGC2018-098153-B-C33, as well as the support of the Generalitat de Catalunya/CERCA programme.  HRAJ acknowledges support from the UK Science and Technology Facilities Council grant number  [ST/M001008/1].  CGT is supported by Australian Research Council grants DP0774000, DP130102695 and DP170103491. JT  was supported by NASA through Hubble Fellowship grant HST-HF2-51399.001 awarded by the Space Telescope Science Institute, which is operated by the Association of Universities for Research in Astronomy, Inc., for NASA, under contract NAS5-26555. GAE is supported by the Ministerio de Ciencia, Innovaci\'on y Universidades Ram\'on y Cajal fellowship RYC-2017-22489 and by the Science and Technology Facilities Council grant number ST/P000592/1 \\

{\noindent\bf{Author Contributions:}}\\
S.V.J led the observing proposal, team coordination, participated in the data analysis and wrote the manuscript\\
S.D. led the data analysis and contributed to the writing of the manuscript \\
J.R.B participated in the writing of the observing proposal, simulations, reviewing manuscript\\
C.A.H participated in the writing of the observing proposal, final consistency checks, and writing the manuscript \\
R.P.N. Contributed the discussion of planetary dynamics and manuscript review \\
E.R. ASH2 photometry, coordination of photometric observations, data analysis \\
M.J.L.G. ASH2 photometry: data reduction\\
N.M. ASH2 photometry: observer\\
R.L, M.Z., S.S.V.,J.J. ran the blind tests, data analysis and manuscript review \\
E.P. data analysis and manuscript review \\
Z.M.B. and M.R.D, Contribution of HARPS data and manuscript review \\
G.A.L.C Contributed to the discussion of planet formation and manuscript review \\
I.R.,  H.R.A.J., A.R., P.J.A, Writing and review of manuscript \\
R.P.B. AAT/PFS data collection and analysis \\
C.G.T., J.B., B.D.C, S.OT.,R.W.,AAT/UCLES observers and review of manuscript \\
J.D.C, F.F., S.A.S., J.T.,  PFS observers \\
G.A.E writing of the observing proposal, data analysis and writing of the manuscript\\

{\noindent\bf{Competing Interests:}} There are no competing interests to declare. \\

{\noindent\bf{Data and materials availability:}} The reduced RVs and photometric data are provided in data S1.
Our HARPS raw data are available in the ESO archive (http://
archive.eso.org) under the program IDs listed in table S1. Reduced
HIRES RVs were taken from (33). The UCLES data are available
from the AAT archive (https://datacentral.org.au/archives/aat/)
by searching the coordinates RA ‘23:05:52h,’ Dec ‘-35:51:11d,’ a
radius of 300 arcseconds, and dates 1998–2012. The PFS spectra,
our dynamical stability simulations, and our Gaussian processes fitting
code are available at https://figshare.com/s/d581c1a17536eeb813ea.
The TESS photometry was retrieved from https://mast.stsci.edu/
portal/Mashup/Clients/Mast/Portal.html, and the All Sky
Automated Survey (ASAS) photometry was retrieved from
www.astrouw.edu.pl/asas/?page=aasc. \\


\begin{table}[p]
\caption{{\bf{Stellar parameters for GJ\,887 and parameters for planets b and c}}. Listed for GJ887 are the parallax in milliarcseconds, distance in parsecs, V-band and GAIA magnitudes, stellar mass as a fraction of the Sun's mass, metallicity relative to the Sun, luminosity and radius in solar units, rotational velocity ($v\sin i$), and surface gravity log g.  The stellar mass was computed using the mass-radius relation of \cite{Schweitzer2019A&A...625A..68S}.  S$_{\mathrm{eff}}$ is the incident flux from GJ887 relative to the incident flux on the Earth from the Sun and T$_{\mathrm{equil}}$ is the equilibrium temperature of the planet. }  
\vspace{0.2cm}
\protect\label{t-stparam}
\centering
\begin{tabular}{l c c | l c  l c }
\hline
\hline
Parameter & Value & Reference & Parameter & GJ\,887\,b & GJ\,887\,c\\
\hline
Spectral type & M1V & \cite{Mann2015ApJ...804...64M}  & $K_p$ [m s$^{-1}$]  & 2.1$^{+0.3}_{-0.2}$ & 2.8$\pm 0.4$   \\
Parallax (mas) & 304.2190 $\pm$ 0.0451 & \cite{GAIA2018A&A...616A...1G}
& $P_p$ [d]  & 9.262 $\pm 0.001$ & 21.789$^{+0.004}_{-0.005}$  \\
Distance (pc) & 3.2871$\pm 0.0005$  & & $m_p$ [$M_\oplus$] & 4.2$\pm 0.6$ & 7.6$\pm 1.2$ 
\\
Magnitude &  $V=$7.34,$G=$6.522 & & $a_p$  [AU] & 0.068$\pm 0.002$ & 0.120$\pm 0.004$ \\
Mass ($M_\odot$) & 0.489 $\pm$ 0.05  &  & $S_{\mathrm{eff},p}$ [$S_{\mathrm{eff},\oplus}$]& 7.95$\pm 0.2$ & 2.56$\pm 0.2$  \\ 
{[}Fe/H{]} & -0.06 $\pm$ 0.08 & \cite{Mann2015ApJ...804...64M} & $T_\mathrm{equil}$ (K) & 468 & 352  \\
$T_\mathrm{eff}$ (K) & 3688 $\pm$ 86 & \cite{Mann2015ApJ...804...64M} \\
Luminosity ($L_\odot$) & 0.0368 $\pm$ 0.004 & \cite{Mann2015ApJ...804...64M}\\
Radius ($R_\odot$) &  0.4712 $\pm$ 0.086  & \cite{Segransan2003A&A...397L...5S,Demory2009A&A...505..205D,Boyajian2012ApJ...757..112B} \\
$v\sin i$ (km s$^{-1}$) & 2.5 $\pm$ 1.0 & \cite{Browning2010AJ....139..504B} \\
$\log R'_{\rm HK}$ mean & -4.805 $\pm$ 0.023& \cite{BoroSaikia2018A&A...616A.108B}\\
log(age/years) & 9.46 $\pm$ 0.58 & \cite{Mann2015ApJ...804...64M}\\
log g & 4.78 & \cite{Rabus2019MNRAS.484.2674R} \\
\hline
\end{tabular}
\end{table}

\newpage

\title{A multiple planet system of super-Earths orbiting the brightest red dwarf star GJ887}


\author
{S.~V.~Jeffers, S.~Dreizler, J.~R.~Barnes, C.~A.~Haswell, ,  R.~P.~Nelson,\and  E.~Rodríguez,  M.~J.~L\'opez-Gonz\'alez, N.~Morales, R.~Luque, \and M.~Zechmeister,  S.~S.~Vogt, J.~S.~Jenkins,   E.~Palle, Z.~M.~Berdi\~nas,  \and  G.~A.~L.~Coleman, M.~R.~D\'iaz, I.~Ribas, H.~R.~A.~Jones,  \and R.~P.~Butler,  C.~G.~Tinney, J.~Bailey,  B.~D.~Carter, S.~O'Toole \and R.~A.~Wittenmyer, J.~D.~Crane,   F.~Feng,  S.~A.~Shectman \and J.~Teske, A.~Reiners, P.~J.~Amado, G.~Anglada-Escud\'e\\
}


\date{}






\maketitle 

\section*{Supplementary Materials}

Materials and Methods\\
Figs. S1 to S11\\
Tables S1 to S7\\
References (34 to 70)

\renewcommand{\thefigure}{S\arabic{figure}}
\setcounter{figure}{0}

\renewcommand{\thetable}{S\arabic{table}}
\setcounter{table}{0}

\renewcommand{\theequation}{S\arabic{equation}}
\setcounter{equation}{0}

\renewcommand{\thesection}{S\arabic{section}}
\setcounter{section}{0}

\section*{Materials and Methods}

\subsection*{Observations and measurements}

In this section we describe the radial velocity and photometric data sets.  

\subsubsection*{Radial velocity time series}

The detection of exoplanetary RV signals requires both a long temporal baseline and dense sampling to identify and robustly characterise long-period signals and sources of correlated noise which can lead to false-positive planet detections.  We use RV observations of GJ~887 covering a baseline of over 20 years. The new RedDots \# 2 observations are clustered towards the end of the dataset and span a time interval of 90 days using a cadence of approximately one observation per clear night.\\

GJ~887 was observed from July to September 2018 using the HARPS spectrograph on the ESO 3.6m telescope at La Silla observatory in Chile as part of the RedDots \#2 program. We obtained 65 observations.  We retrieved from the HARPS archive an additional 72 observations taken between December 2003 and December 2017.  All HARPS data were wavelength calibrated using a hollow-cathode lamp and extracted and calibrated using the HARPS Data Reduction Software (DRS) \cite{Mayor2003Msngr.114...20M,Lovis2007A&A...468.1115L}. The Doppler shift measurements were made with the Template Enhanced Radial velocity Application software (TERRA) \cite{Anglada2012ApJS..200...15A}. 
For analysis, the HARPS data were divided into two periods: before and after the fiber change in May 2015 \cite{locurto2015Msngr.162....9L} as this could affect the line-spread function. Stitching effects were corrected using TERRA.  We used  151  archival observations (see Table~\ref{tab:data_properties}). These were from (i) the HIRES spectrograph mounted on the Keck I 10-m telescope located on Mauna Kea, Hawaii from June 1998 to December 2013; (ii) the PFS spectrograph at the Magellan II 6.5-m telescope at Las Campanas Observatory in Chile from August 2011 to November 2013; and (iii) the UCLES spectrograph located at the 3.9\,m Australian Astronomical Telescope at Siding Springs Observatory from August 1998 to July 2012. These three spectrographs use iodine cells for stable wavelength calibration \cite{Ribas2018Natur.563..365R}.  The HIRES data have been corrected for nightly zero point systematics\cite{Tal-Or2019MNRAS.484L...8T}.

\subsubsection*{Photometric time series}

Photometric data, monitoring intrinsic stellar brightness variations from e.g. stellar activity such as starspots and the rotation period of the star, are listed in Table \ref{tab:phot-obslog}. {{B}} band observations were made from July to October 2018 at San Pedro de Atacama Celestial Explorations Observatory (SPACEOBS) in Chile using the 40 cm robotic telescope ASH2 \cite{Sicardy2011Natur.478..493S}. These observations were taken almost simultaneously with the RedDots \# 2 HARPS observations.  More than 150 additional {{V}} band observations from a more than two year time-span during 2002-2004 were incorporated from the archival survey ASAS (All-Sky Automated Survey \cite{Pojmanski1997AcA....47..467P}). GJ~887 was also observed over a time span of 27.4 days in autumn 2018 during Sector 2 of the TESS space survey.\\

\subsubsection*{Stellar activity time series}

Magnetic activity on the surface of the star can induce an additional apparent RV signal that can lead to a false-positive planet.   Spectral lines that are known to be sensitive to the star's magnetic activity can trace different aspects of this activity.  We extract a time-series of measurements for commonly used stellar activity indices and spectral lines that are known to be sensitive to the star's magnetic activity: the H$\alpha$, H$\beta$ and Na D spectral line fluxes and the S-index.  The S-index is computed using \cite{Duncan1991ApJS...76..383D} 
\begin{equation}
{\rm S} = (H + K)/(V +R)
\end{equation}
where the values for $H$ and $K$ are fluxes at the line cores, using triangular pass-bands, and $V$ and $R$ are the nearby continuum regions as listed in Table~\ref{tab:mag-activity}.  The other indices are computed following established methods \cite{Barnes2016MNRAS.462.1012B} with the central wavelengths, the pass-band widths, and the associated continuum regions specified in Table~\ref{tab:mag-activity}. \\

\subsection*{Analysis of time-series}

For the analysis of the time-series data we first search for potential signals using a log-likelihood periodogram.  We then apply global fits with a more sophisticated model using Gaussian processes that also incorporates correlated noise such as that originating from stellar magnetic activity.\\

\subsubsection*{Model of the data and significance assessment}

We analyze the data using a Doppler model, and use a statistical figure-of-merit tool to assess the goodness of fit of the model to the data. These tools are identical to previous studies \cite{Ribas2018Natur.563..365R}, so we only briefly summarize them here. The Doppler model describes the radial velocity $v$ and properties of the star and planet as they orbit a common center of mass.  For each observation $i$ at time $t_i$, the velocity can be described as: 
\begin{equation}
    v(t_i) = \gamma_{\mathrm{INS}} + S\cdot (t_i-t_0)
+ \sum_{p=1}^{n} v_p(t_i)
\end{equation}
where the free parameters are $\gamma_{\mathrm{INS}}$, a constant offset for each instrument, and $S$, a linear trend. $t_0$ is the time at periastron passage,  and $v_p$ is the planet's velocity 
\begin{equation}
    v_p(t_i) = K_p \cos \left[ \nu_p(t_i,P_p,t_{0,p},e_{p})+\omega_p \right] + e_p \cos \bar{\omega}_p\,\,,
\end{equation}
where $K_p$ is the Doppler semi-amplitude of the planet $p$, $P_p$ is the orbital period, 
$e_p$ is the orbital eccentricity,  $\omega_p$ is the argument of periastron of the orbit, and $\nu_p$ is the function for the true anomaly \cite{Lucy2005A&A...439..663L}.  In the case of circular orbits, this equation becomes
\begin{equation}
    v_{p,\mathrm{circ}}(t_i) = K_p \cos \frac{2\pi(t_i-t_0)}{P_p}\,\,.
\end{equation}

\noindent When analysing time-series for the stellar activity indices and photometry  we also assume this pure sinusoidal model for computational efficiency and simplicity in the interpretation of the signals, but the procedure is otherwise identical. \\ 

The goodness of fit of the model ($v_i$) to the data is quantified by maximising the likelihood function $L$.  For measurements with normally distributed noise, $L$ can be written as
\begin{eqnarray}
L &=& \frac{1}{(2\pi)^{N_{obs}/2}} |C|^{-1/2} 
\exp{\left[-\frac{1}{2}
\sum_{i=1}^{N_{obs}} \sum_{j=1}^{N_{obs}} 
r_i r_j C_{ij}^{-1}
\right]}\,\,, \label{eq:likelihood}\\
   r_i &=& v_i - v(t_i) \, \,,
\end{eqnarray}
where $r_i$ is the residual of each observation $i$, N$_{obs}$ are the number of observations, $C_{ij}$ are the components of the covariance matrix between measurements $i$ and $j$, and $|C|$ is its determinant. This model incorporates simultaneous modelling of intrinsic stellar variability using Gaussian processes (see below for details). We use a frequentist False Alarm Probability of detection (FAP) as a statistical test of the significance of a new signal\cite{Baluev2009MNRAS.393..969B}; where we use FAP\, $ < 10^{-3}$ ($0.1\%$) as our detection threshold. \\

\subsubsection*{Analyses of time-series : RV data}

\begin{figure*}
    \centering
    \includegraphics[width=\textwidth]{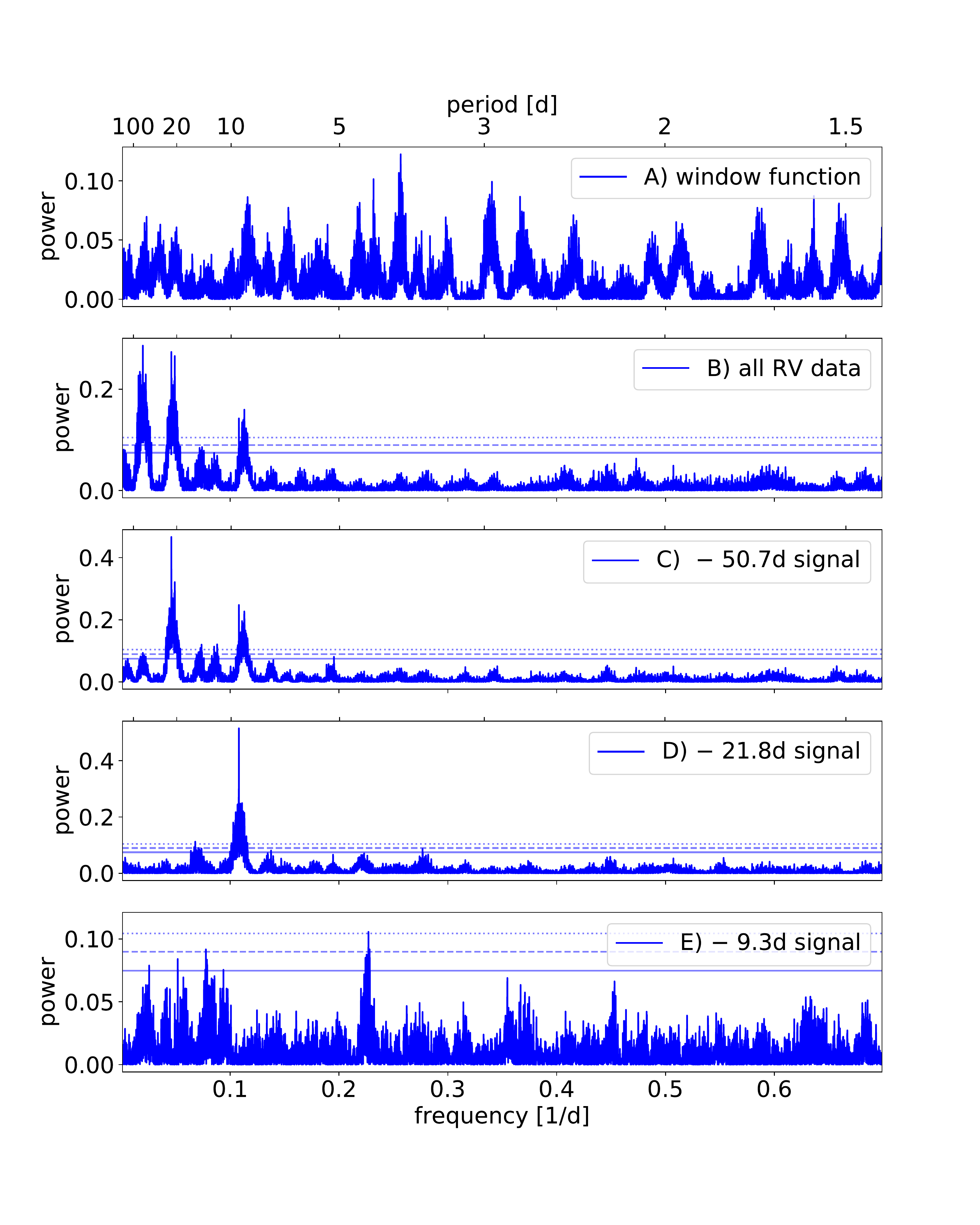}
    \vspace{-2.0cm}
    \caption{\label{fig_prewhite} {\bf{Periodogram search of signals in the RV data}}. From Panels A to E: The window function (panel A), identification of the first signal (50.7 days, panel B), after removal, search for the second signal (21.8 days, panel C), after removal, identification of the third signal (9.3 days, panel D), and final periodogram with no more signals left. The solid, dashes and dotted lines indicate 10\%, 1\%, and 0.1\% False Alarm Probability levels.}
    \end{figure*}

We perform the initial signal search using log-likelihood periodograms with Keplerian (RV) or sinusoidal signals (RV and activity proxies). For computational efficiency, this initial periodogram signal search assumes uncorrelated measurements (that is, the covariance matrix $C_{ij}$ in equation (\ref{eq:likelihood}) is assumed diagonal (i.e. it is defined as $C_{ij} = \left(\epsilon_i^2 + s^2_{INS}\right)\delta_{ij}$, which is equivalent to assuming uncorrelated measurements or white noise). Detection periodograms are shown in Fig.~\ref{fig_prewhite}, and the values of the improvement in the log-likelihood statistic for the different models with 0, 1, 2 and 3 signals are presented in \ref{tab:modelcomparison}.\\

For the RV data, this first signal identification is then re-evaluated by using a more complete model including more general parameterizations of the covariances using Gaussian processes. We use the solutions found in the periodograms as the initial values for numerical optimization routines in {\sc{scipy.optimize}} \cite{2020SciPy-NMeth} to converge to the local maximum likelihood model, followed by a Monte Carlo Markov Chain sampling of the posterior solutions using {\sc emcee \cite{Foreman-Mackey2013PASP..125..306F}} with 400 walkers and 20000 steps. The chains are initialized with a Gaussian distribution using the preliminary values coming form the periodograms, and 1000 times the standard deviation from the likelihood minimization. This initialization is far broader than the final posterior distribution and ensures that the parameter space gets sufficiently explored. Boundaries for the parameters are only set where a positive definite value is physically required, for example for the orbital period.\\

To ascertain the significance of a signal, we optimize the likelihood of a model without the investigated Keplerian orbit as a baseline. This model includes all of its other components such as correlated noise model, offests, jitters and other Keplerians.   We then compare it to the maximum likelihood of the same model with the new signal. The improvement of the likelihood statistic $\Delta ln L$ is then used to make a FAP assessment \cite{Baluev2009MNRAS.393..969B} .\\

The correlated noise results from intrinsic covariances in the measurements.  We model the correlated noise using Gaussian Processes as provided by the {\sc celerite} package \cite{2017AJ....154..220F}. The kernels used to parameterise the covariances were a damped exponential kernel (REAL), and a SHO kernel (stochastically excited harmonic oscillator). The REAL kernel only contains two free parameters (amplitude $a$ and decay time-scale $\tau$), to model covariances that decay exponentially over time. The SHO kernel also contains an amplitude and a time-scale, but it also has the period of the corresponding harmonic oscillator as one additional free parameter. If the signature of stellar rotation is present in the data, the SHO kernel typically provides a better fit that the REAL kernel.\\

Including the REAL kernel to model correlated noise results in a significant improvement compared to the two planet model without Gaussian processes ($\Delta \ln L= + 62$, see Table~\ref{tab:modelcomparison}). However, and despite having more flexibility, the SHO kernel leads to a similar maximum likelihood value as the REAL kernel, indicating that there is no clear signal of stellar rotation. Also, when using REAL kernel the addition of a third signal (at 50.7-d) does not improve the likelihood statistic significantly. Moreover, when running an MCMC starting at the nominal three planet solution, the amplitudes and periods for the third signal become unconstrained. As a result, we conclude that a third planet with a signal of $\sim$50\,d is not supported by the current RV dataset.\\

\begin{figure}
    \centering
    \includegraphics[width=\textwidth]{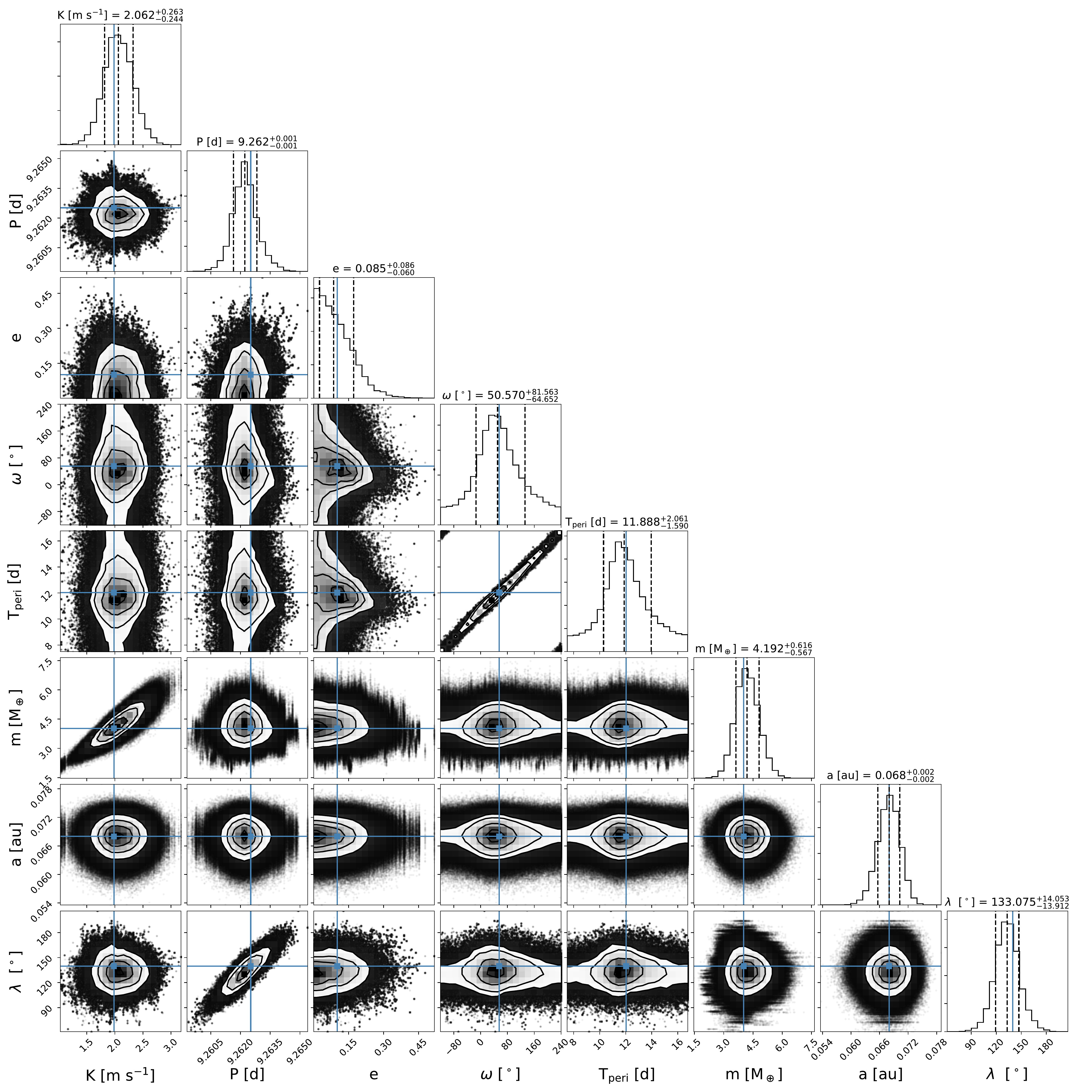}
    \caption{{\bf{Parameter distributions for planet GJ\,887\,b from the two planet and REAL noise kernel fit}}. The diagonal shows the posterior distribution of each parameter, the off-diagonal plots show the two parameter correlations for all combinations. Contour lines show the 0.5, 1, 1.5, and 2 $\sigma$ levels. The best fit values for the parameters are indicated using the horizontal and vertical solid blue lines. The vertical dashed lines on the histogram plots show  the 16\%, 50\%, and 84\% percentiles.}
    \label{fig:cornerPlanetb}
\end{figure}

\begin{figure}
    \centering
    \includegraphics[width=\textwidth]{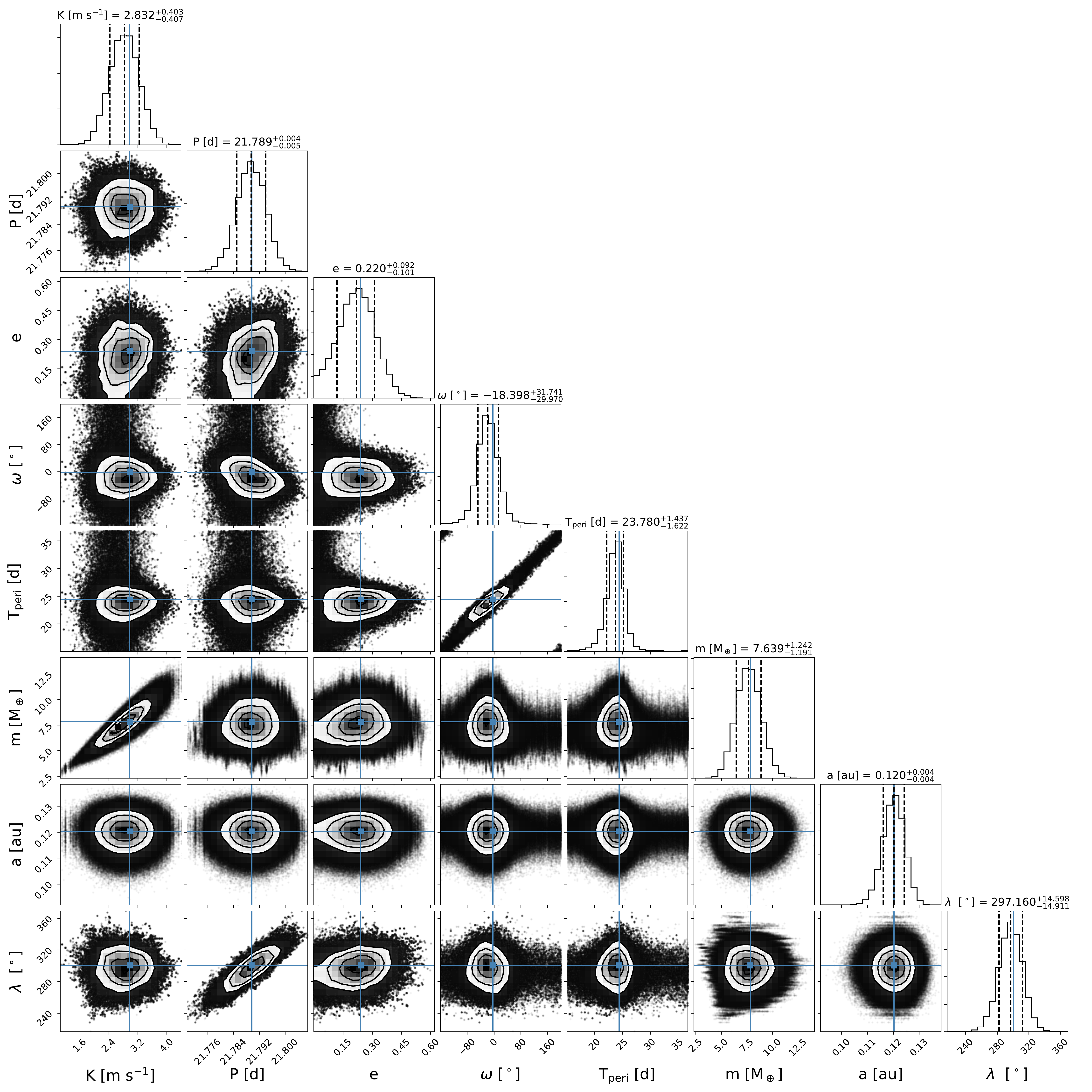}
    \caption{{\bf{As Figure S4 but for planet GJ\,887\,c.}}}
    \label{fig:cornerPlanetc}
\end{figure}

\begin{figure}
    \centering
    \includegraphics{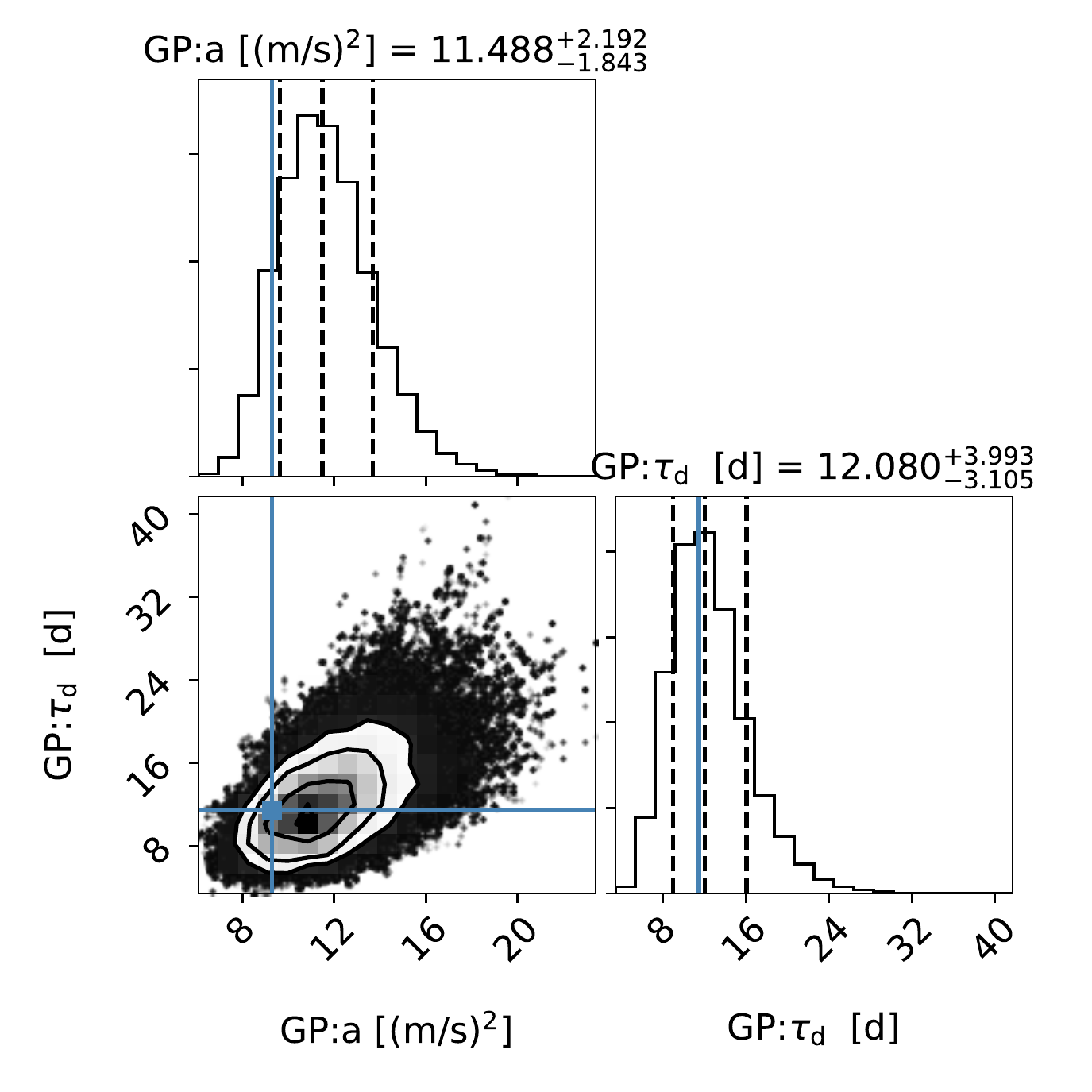}
    \caption{{\bf{As Figure S4 but for the hyper parameters of the Gaussian Processes REAL model.}}}
    \label{fig:cornerGP}
\end{figure}

The detection sequences for models with increasing complexity are listed in \ref{tab:modelcomparison}, and the best fit parameters for the reference model (2 Keplerians with the REAL Gaussian processes kernel) are presented in the main manuscript (Table~\ref{t-stparam}).  While  $K$,  $P$, $e$ as well as $\omega$ and $t_0$ are direct fit parameters, the semi major axis $a$, the planetary minimum mass $m$, and the mean longitude $\lambda$, are derived ones, and can depend on the value of astrophysical quantities with uncertainties. For a realistic estimation on the uncertainties in the semi-major axis and the minimum mass, we draw samples from the MCMC distributions for the fitted parameters, and assume a normal distribution for the stellar mass with mean $0.489$\,M$_\odot$ and standard deviation $\pm 0.05$\,M$_\odot$. The priors for the fit are listed in Table\,\ref{tab:priors}. The posterior distributions with the median value as well as the 16$\%$ as well as $84\%$ percentile are displayed in Fig.\,\ref{fig:cornerPlanetb}, \ref{fig:cornerPlanetc}, and \ref{fig:cornerGP}.  The values of the individual instrumental offsets and jitter parameters are given in Table~\ref{tab:jitteroffset}. \\

As an additional experiment, we also explored fitting an SHO kernel to the time-series of the ASAS photometry and the S-index in an attempt to determine the rotation period of the star using the time-series of the activity proxies.  As with the RV data, no oscillator period could be determined using the SHO kernel, where the MCMC failed to converge to a precise value. This indicates that the lifespans of active regions could be shorter than the stellar rotation period, which remains unknown.\\

We detect robust signals at periods of 9.2 and 21.3~d in the RV data. Correlated noise seems strongly present, and has a correlation decay time-scale $\tau$ of $\sim 12$ d (99\% credibility interval between 7 and 24-d, see Fig.~\ref{fig:cornerGP}). However, the fits using an SHO kernel do not converge to any particular time-scale for stellar rotation. Since correlations seems to explain most of the RV variability, there is not enough support for a third Keplerian signal in the current RV dataset. \\

\subsubsection*{Analyses of time-series : stellar activity indicies}

\begin{figure}
    \centering
    \includegraphics[angle=270,width=0.8\textwidth]{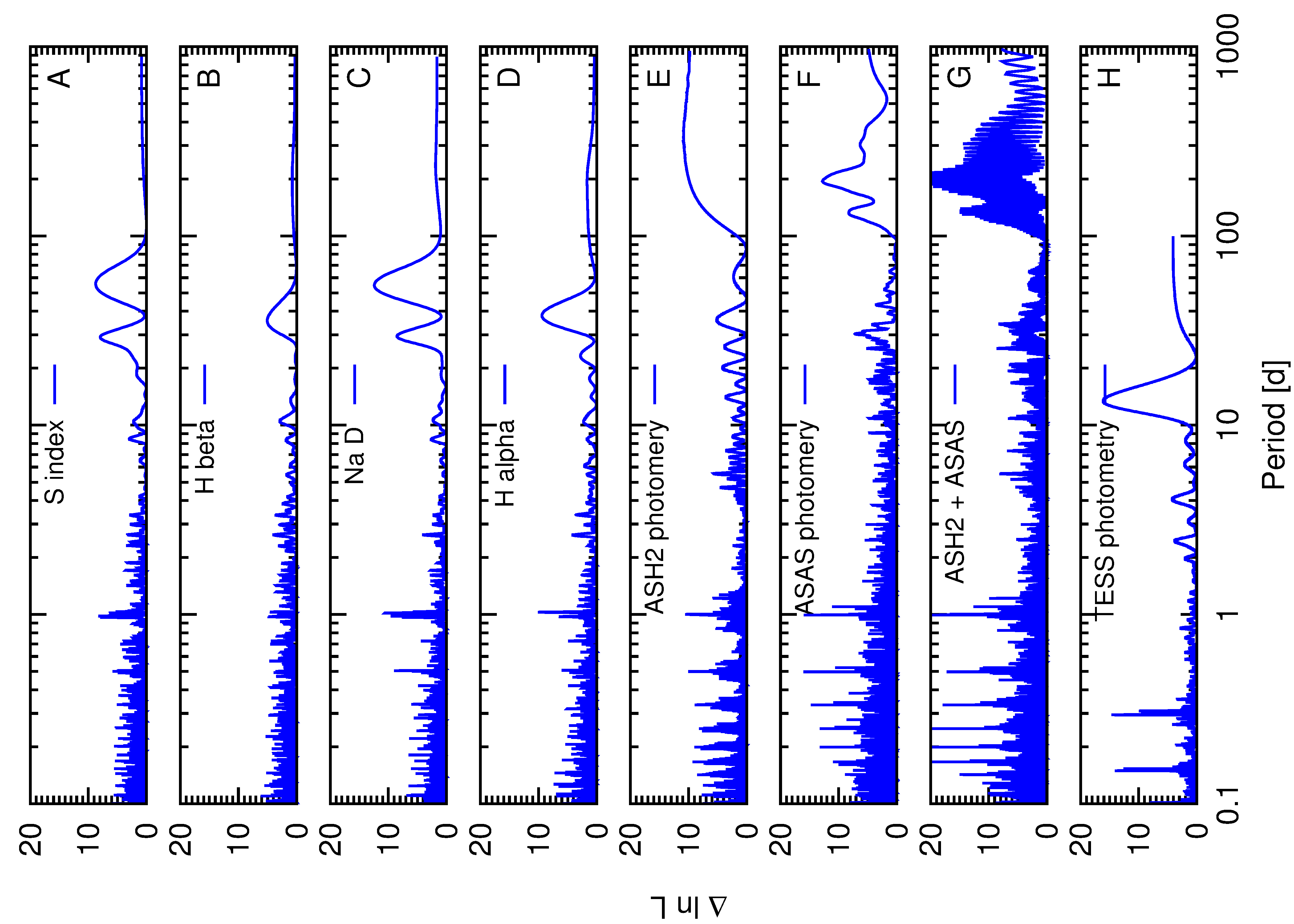}
    \caption{{\bf{Periodograms for the stellar activity indicies and photometric data}}.  The stellar activity indicies are shown in panels (A) to (D) and the photometric data is shown in panels (E) to (H). The corresponding periods are tabulated in Table ~\ref{tab:mag-activity}. Apparent periodicities at $\leq$ 1 day are spurious.}
    \label{fig:periodograms}
\end{figure}

\begin{figure}
    \centering
    \includegraphics[angle=270,width=0.8\textwidth]{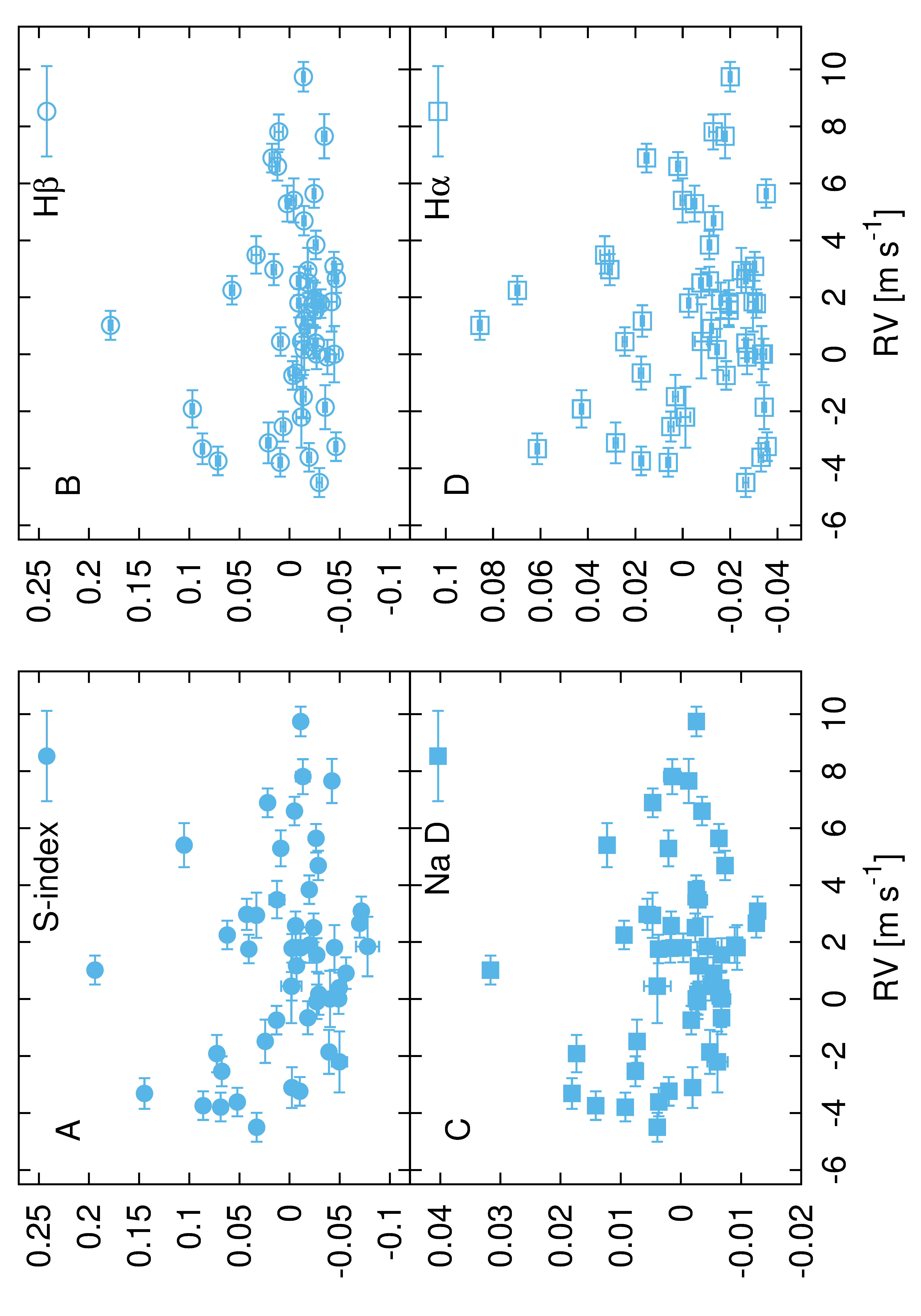}
    \caption{{\bf{Scatter diagrams of activity indices with RV}}.  Simultaneous measurements of RV versus panel A: the S-index; panel B: H$\beta$; panel C: NaD; panel D: H$\alpha$.}
    \label{fig:Rv-activity}
\end{figure}

The Red Dots \# 2 observations have continuous coverage of GJ\,887 for 90 nights.  We searched for correlations between the activity indicators and with the RVs derived for this time series.  The results are listed in Table~\ref{tab:activity_corr} and the corresponding periodograms are shown in Fig.~\ref{fig:periodograms}. 
We find the strongest correlation between H$\alpha$ and H$\beta$ with a Pearson's correlation coefficient of $r=0.89$ and a Student's t-test probability (stp) $= 2.06\times$10$^{-21}$ \cite{Press1996}. We also find weak anti-correlations (with r $<$ 0.3) between the activity indicies and the RV as shown in Fig.~\ref{fig:Rv-activity}.
However, values of stp $>$ 0.05 imply no strong evidence to reject the null hypothesis of no correlation. We find potential periodicities in the S-index and Na\,D with a period of approximately 54.9 and 55.8 days, respectively, while there is a potential period of 37.9 days in the H$\alpha$ and 35.5 days in the H$\beta$ spectral lines (Table~\ref{tab:mag-activity} and Fig.~\ref{fig:periodograms}).  The discrepancy in the derived periods for S-index and Na\,D compared to H$\alpha$ and H$\beta$ could reflect different timescales for activity on GJ\,887.  However, the time span of our observations is too short to determine the reason for differing periods.\\

\begin{figure}
    \centering
    \includegraphics[width=\textwidth]{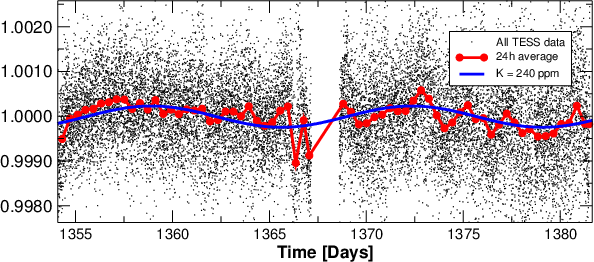}
    \caption{{\bf{TESS photometry of GJ~887}}. The black dots are the detrended TESS observations obtained by the mission pipeline (so called Pre-search Data Conditioning Simple Aperture Photometry flux). The red points are 24h averages of the same data. The blue line is a possible sinusoidal periodicity extracted from the 24h averaged observations which has a semi-amplitude of 240 ppm and a period of ~13.7 days.  We advise caution in interpreting  this low amplitude periodicity as the stellar rotation period because it could result from instrumental systematics.}
    \label{fig:tess-lc}
\end{figure}

In the combined photometric data (ASH2+ASAS) set we find a period of approximately 200 days with a $\Delta$ $\ln L$ value of 22.5.  The individual data sets are listed in Table~\ref{tab:mag-activity}.  The residuals show possible further signals at periods between 30-60\,d.  All periods are candidates for rotation, though the longer 200 day rotation period is unlikely for star with a mass of 0.49M$_\odot$ \cite{McQuillan2014ApJS..211...24M,Newton2018AJ....156..217N} as a typical rotation period is about 60 days. The TESS observations show smooth variability in the photometry of GJ~887 with a semi-amplitude of about 240 ppm semi-amplitude (or 480 ppm peak-to-peak).  In Fig.~\ref{fig:tess-lc} we show the TESS Pre-search Data Conditioning Simple Aperture Photometry flux (PDCSAP) pipeline light curve and 24 hour averages where the potential periodicity with a period of 13.7 days with a semi-amplitude of 240 ppm is shown. We regard this value as an upper limit as such a low amplitude periodicity may not be the stellar rotation period as systematic errors on the order of a few days in the TESS photometry might be dominating the signal. No other signals are present above 100 ppm.\\

GJ\,887's log($R^{'}_{HK}$)= -4.805 \cite{BoroSaikia2018A&A...616A.108B} implies a rotation period of between 10 and 60 days \cite{AstudilloDefru2017A&A...600A..13A}.  However, GJ~887 does not show a distinct peak in this period range in neither the photometry nor activity indices.   The inferred rotation period using log($R^{'}_{HK}$) is based on stars with significantly higher magnetic activity levels, and consequently a greater starspot coverage which shows a well defined rotational modulation.  \\

Even with the extensive photometric data set we cannot confirm a rotation period of the order of a few tens of days, or exclude the possibility that very inactive stars such as GJ\,887 could have much slower rotation. This is consistent with previous studies where only 10\% of early M dwarfs such as GJ 887 show detectable rotation periods \cite{Browning2010AJ....139..504B}.\\

\subsubsection*{Analyses of time-series : Additional blind tests on RV data}

As an additional check on the statistical significance of the signals, and to avoid any confirmation biases, as the archival data already showed evidence of several signals, we distributed the time-series among several sub-teams within the RedDots collaboration. No prior information on the possible signals was provided to these sub-teams.  Here we provide a summary of the different approaches and conclusions drawn from the experiment. The four independent methods/sub-teams were : \#1 the  Exo-Striker tool
\#2 the Exoplanet Mcmc Parallel tEmpering Radial velOcity fitteR ({\sc{EMPEROR}}; \cite{Jenkins2019MNRAS.487..268J}) \#3 Systemic \cite{2009PASP..121.1016M} and \#4 Juliet \cite{juliet2019MNRAS.tmp.2366E} codes to analyse the radial velocity data for GJ~887.\\

\noindent\underline{Method \#1}  We employed the Exo-Striker tool \cite{Trifonov2019ascl.soft06004T} on the five RV data sets. Using prewhitening with the generalised Lomb-Scargle periodogram \cite{Zechmeister2009A&A...496..577Z}, there are three significant signals with periods of 22 days (${\rm FAP}=3\cdot 10^{-16}$), 9 days (${\rm FAP}=3\cdot 10^{-9}$) and 51 days (${\rm FAP}=1\cdot 10^{-11}$).  A final simultaneous fit with three Keplerians and jitter results in moderate eccentricity parameters and changes of the amplitudes. \\

\noindent\underline{Method \#2} {\sc{emperor}} uses Markov chain Monte Carlo samplings, coupled with Bayesian statistics, to probe the multi-dimensional posterior probability distribution.  It makes use of the EMCEE sampler \cite{Foreman-Mackey2013PASP..125..306F} in parallel-tempering mode to ensure that the highly multi-modal posterior is well sampled. We employ {\sc{emperor}} in the default automatic mode, and begin by analysing the data using a flat noise model, providing baseline statistics which allow the code to determine if any subsequent signal is statistically significant.  After running the base noise model, a single Keplerian signal is introduced, returning a detection that has a period of approximately 22 days. We then ran {\sc{emperor}} with a $k=2$ model, detecting another signal with a period of approx 9 days.  Finally, a third Keplerian is detected with a period of 51 days. The {\sc{emperor}} results show three statistically significant signals present in the data.\\

\noindent\underline{Method \#3} The {\sc systemic} models were all simple summed Keplerians, without invoking any planet-planet dynamical interaction. Parameter values and their uncertainties (standard deviation) are averages from a 1000-iteration bootstrap run. 
The 22\,d and 9\,d signals are well-fit as summed Keplerians. The 51\,d signal appears in the residuals of the 2-planet model.  It is substantially broader than the first two signals and has the shape and breadth of a signal produced by stellar activity and / or stellar rotation. \\

\noindent\underline{Method \#4} The {\sc juliet} models have been described previously by \cite{Luque2019A&A...628A..39L}.  For GJ\,887, models were run using a combination of 2 and 3 signals both with and without Gaussian processes.  The {\sc juliet} models detect two planets orbiting at periods of 9.26 days and 21.7 days.  A simple exponential Gaussian processes  kernel can  account for the correlated noise especially in the 30-60d range. A simple Keplerian cannot model the periodicity at $\sim$50\,d. \\

All three RV signals were detected and reported independently by the sub-teams. Two of the sub-teams (Methods \#3 and \#4) independently concluded that the correspondence of the third signal to a true Keplerian, or exoplanet orbit, is questionable and is consistent with the more detailed analysis presented in this paper.
  
\section*{Planetary system architecture and dynamical considerations}

\subsection*{GJ~887 in the planetary system architecture context}
\begin{figure}
    \centering
    \includegraphics[angle=270,width=\textwidth]{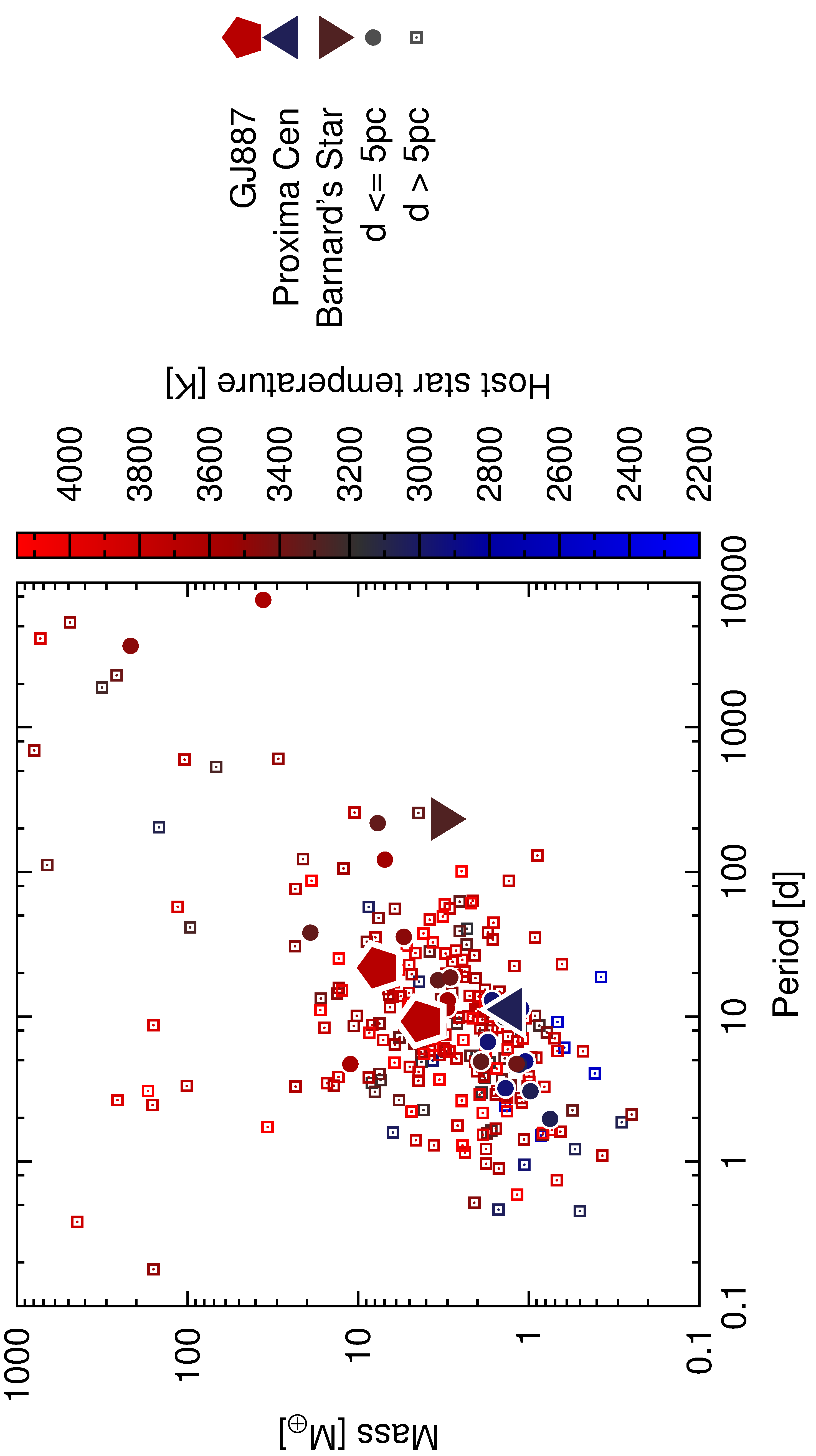}
    \caption{{\bf{Minimum planet mass as a function of orbital period for all known planets orbiting M dwarfs}}.  We use the mass to radius relation of  \cite{Weiss2013ApJ...768...14W}.   Colours indicate host star effective temperature, see colour bar. The two large red pentagons indicate GJ~887\,b and GJ~887\,c.}
    \label{fig:P-M-RedDots}
\end{figure}

\begin{figure}
    \centering
    \includegraphics[angle=270,width=0.9\textwidth]{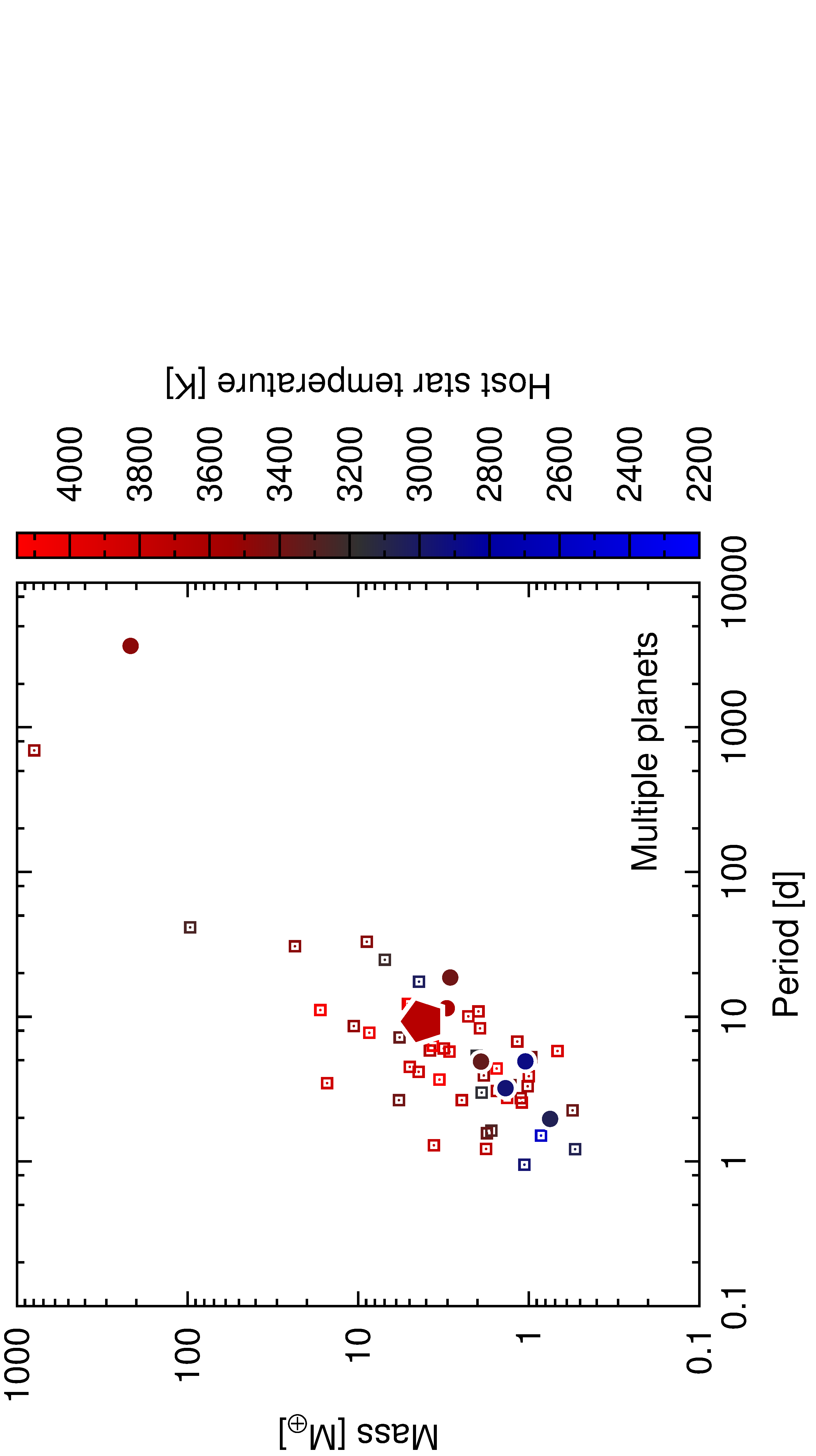}
    \caption{{\bf{The innermost known planet for known M dwarf multi-planet systems.}} As for Fig.~\ref{fig:P-M-RedDots}. The innermost planet of GJ~887 is comparatively long period compared to other multi-planet systems.}
    \label{fig:P-M-single_multiple}
\end{figure}

In Fig.\,\ref{fig:P-M-RedDots} GJ\,887\,b and c are shown in the orbital period -- planet mass plane together with all known planets orbiting M dwarfs.  GJ\,887\,b and c appear fairly typical, but are towards the top of the mass distribution and orbit the brightest M-dwarf. This is consistent with evidence from the Kepler Mission that masses of super-Earth planets increase with the mass of the host star \cite{Wu2019ApJ...874...91W}
In Fig.\,\ref{fig:P-M-single_multiple} the innermost known planet of the known M dwarf multiple planetary system are shown. GJ\,887\,b is at the long orbital period end of this distribution, and is  
relatively massive for the innermost planet in a multiple system.  Our results and other investigations \cite{2019hsax.conf..266B} have failed to detect shorter period planets than GJ 887~b, and also rule out that any of the signals reported here are caused by aliasing of sub-day period signals.\\

\subsection*{Planetary system stability}

For systems of two or more planets, there are no generally applicable analytical criteria that can be used to determine the long-term stability of the system. In the limiting case of two planets on circular orbits, a system is said to be Hill stable (i.e. the orbits of the planets cannot cross one another) if the following criterion is satisfied \cite{Gladman1993Icar..106..247G}:

\begin{equation}
D_{bc} \equiv \frac{a_{\rm b} -a_{\rm c}}{R_{\rm H}} \ge 2 \sqrt{3},
\label{eqn:Dbc}
\end{equation}
where $a_{\rm b}$ and $a_{\rm c}$ are the semi-major axes of the outer and inner planets, respectively, and $R_{\rm H}$ is the mutual Hill radius defined by
\begin{equation}
R_{\rm H} = \frac{a_{\rm b}+a_{\rm c}}{2} \left(\frac{\mu_{\rm b} + \mu_{\rm c}}{3}\right)^{1/3},
\label{eqn:RH}
\end{equation}
where $\mu_{\rm b}= m_{\rm b} /M_*$, $\mu_{\rm c}= m_{\rm c}/M_*$, $m_{\rm b}$ and $m_{\rm c}$ are the masses of the inner and outer planets, respectively, and $M_*$ is the mass of the central star. The preferred solution for the GJ~887 system obtained for two planets on Keplerian orbits with the REAL Gaussian process kernel (see Table~\ref{t-stparam} in main text) yields semi-major axes $a_{\rm b}=0.068$~au and $a_{\rm c}=0.12$~au, so for circular orbits $D_{bc}\sim 17$ and the system is Hill stable, in agreement with our {\sc{mercury6}} simulations. The preferred solution for two planets, however, yields eccentricities of $e_{\rm b}=0.09^{+0.09}_{-0.06}$ and $e_{\rm c} = 0.22^{+0.09}_{-0.10}$, respectively, and a two planet system with eccentric orbits is Hill stable if the following criterion is satisfied \cite{2013MNRAS.436.3547G}
\begin{equation}
\left(\mu_{\rm b} + \mu_{\rm c} \frac{a_{\rm b}}{a_{\rm c}}\right) 
\left( \mu_{\rm b} \gamma_{\rm b} + \mu_{\rm c} \gamma_{\rm c} \sqrt{\frac{a_{\rm c}}{a_{\rm b}}}\right)^2 > \alpha^3 + 3^{4/3} \mu_{\rm b} \mu_{\rm c} \alpha^{5/3},
\label{eqn:Hill_ecc}
\end{equation}
where $\gamma_{\rm b}=\sqrt{1-e_{\rm b}^2}$, $\gamma_{\rm c}=\sqrt{1-e_{\rm c}^2}$ and $\alpha=\mu_{\rm b} + \mu_{\rm c}$. The two planet solution satisfies the Hill stability criterion \ref{eqn:Hill_ecc} if we adopt the nominal values $e_{\rm b}=0.09$ and $e_{\rm c} = 0.22$, but marginally fails the criterion if we adopt the maximum eccentricities allowed by the quoted uncertainties. Our {\sc{mercury6}} simulations of two planet systems were found to be stable for all values of the eccentricities, a result that is consistent with previous numerical studies of planetary system stability \cite{2013MNRAS.436.3547G}, which show the region of Hill stability is approximately 10\% larger than indicated by \ref{eqn:Hill_ecc}.  It is possible there is a third planet in the GJ~887 system, and the stability criteria \ref{eqn:Dbc} and \ref{eqn:Hill_ecc} are not applicable in that case. Instead we need to consider the AMD stability of the system.

\subsubsection*{AMD stability}
The angular momentum deficit (AMD) of a planetary system containing $N$ planets is defined by \cite{2017A&A...605A..72L}
\begin{equation}
{\cal C} = \sum_{k=1}^N \Lambda_k \left( 1 - \sqrt{1-e_k^2} \cos{i_k} \right),
\label{eqn:AMD}
\end{equation}
where $\Lambda_k = m_k \sqrt{G M_* a_k}$. The AMD is the difference between the angular momentum that the system would have if the planets were on circular orbits, versus the angular momentum it has with the planets possessing eccentricities $e_k$ and inclinations $i_k$ about the invariable plane. For a system where changes occur on long time scales, and mutual perturbations associated with mean motion resonances and those which occur on short time scales are ignored, such that the secular approximation can be used, the semi-major axes of the planets are conserved. In such a system the total AMD is also conserved, and the concept of AMD stability can be applied.\\

\begin{figure*}
\includegraphics[width=4.9cm]{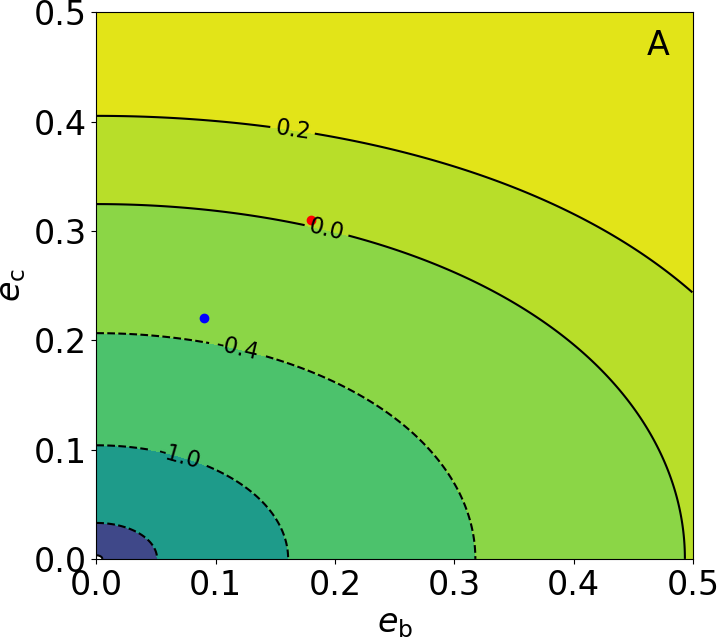}
\includegraphics[width=4.9cm]{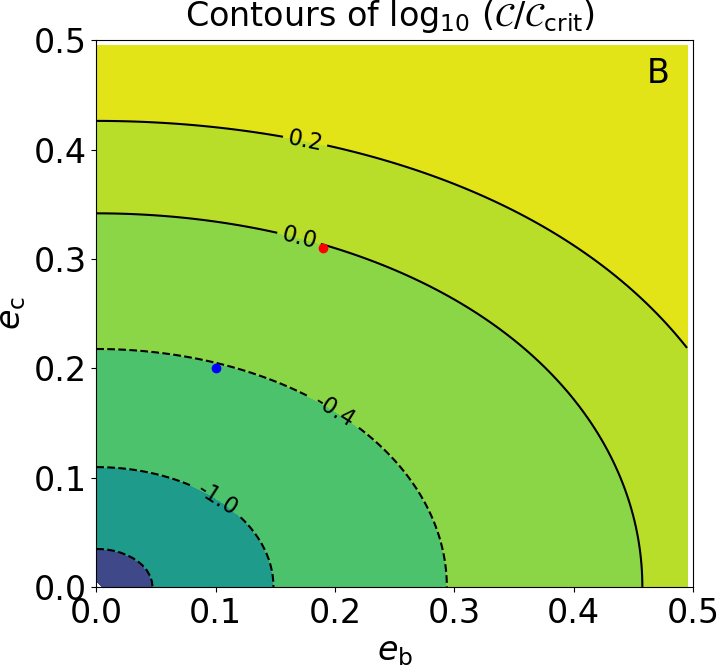}
\includegraphics[height=4.4cm,width=5.5cm]{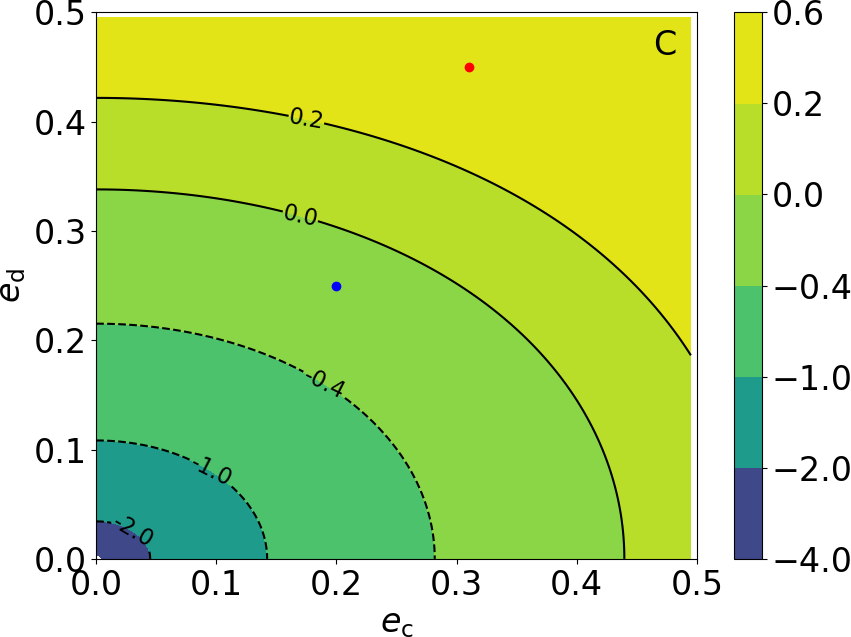}
\caption{{\bf{
Contour plots showing the logarithm of the ratio of the AMD to its critical value for pairs of planets.}} (A): Results for the two planet solution obtained using the REAL Gaussian processes kernel. (B) and (C): Results for the inner and outer pairs of planets, respectively, obtained from the 3 planet Keplerian solution. A system is AMD stable if $\log_{10}{({\cal C}/{\cal C_{\rm crit}})}<0$. The dotted contours show AMD stable regions, and the solid contours show AMD unstable regions. The blue dots show the eccentricity values of the nominal solutions, and the red dots show the upper limits set by the MCMC runs for the values of the eccentricities.}
\label{fig:AMDContours}
\end{figure*}

We now consider the AMD stability of the GJ~887 system \cite[their equations 28, 29, 35 and 39]{2017A&A...605A..72L}. Assessing the stability of a system containing $N>2$ planets involves examining the AMD of each planet pair. We begin by considering the reference solution with 2 planets and the the REAL Kernel. We assume the planetary orbits are coplanar, and we take the masses and semi-major axes to have fixed values corresponding to the nominal fit values in Table~\ref{t-stparam} of the main manuscript. The AMD stability then just depends on the eccentricities. Fig.~\ref{fig:AMDContours} shows contours of $\log_{10}{({\cal C}/{\cal C_{\rm crit}})}$, where ${\cal C_{\rm crit}}$ is the critical AMD that allows the two planet orbits to just intersect, and hence defines the transition to instability. We find that the favoured two planet solution is stable, and only the maximum allowed eccentricities lead to an unstable system.\\

We now consider the AMD stability of the 3 planet Keplerian solution. The results are shown in Figure~\ref{fig:AMDContours}. The nominal 3 planet Keplerian solution is stable, but the outer pair is close to AMD instability, and even with only moderate increases in the eccentricities the system is AMD unstable. If the system had the maximum allowed eccentricities then it would be unstable.\\

\subsubsection*{Hill stability in $N>2$ planetary systems}

The above discussion of AMD stability applies only to systems which evolve according to the secular approximation, where the AMD is conserved. In close packed systems high frequency perturbations influence planetary orbits, and mean motion resonances can play a role.  In these cases, the stability of a general planetary system with $N >2$ planets can only be demonstrated using direct numerical simulations. There have been numerous studies of this problem for planets on initially circular orbits, and with constant spacing between the planets in terms of the mutual Hill radius, $R_{\rm H}$ \cite{1996Icar..119..261C,2002Icar..156..570M}. These studies have allowed scaling relations to be derived that give the typical stability life time of a system in terms of the mutual separations between the planets. The effects of eccentricity and mutual inclination have been considered on the dynamical stability of planetary systems consisting of super-Earths \cite{2015ApJ...807...44P}, for planet masses in the range $3 \le m_{\rm p} \le 9$~M$_{\oplus}$ orbiting a solar mass star, and systems of 7 planets. As such, the results are not directly applicable to the GJ~887 system, but provide a guide to what we should expect. \\

Simulation of planets on initially circular orbits show that the median life time of a system before instability sets in depends on the separation between planets (expressed in units of the mutual Hill sphere). The stability can be expressed in terms of $D_{\rm 50}(t')$, the separation required between planet pairs for 50\% of systems to survive for time $t$ where $t'=t/T_1$, and $T_1$ is the orbital period of the innermost planet in the system:
\begin{equation}
D_{\rm 50}(t') \approx 0.7 \log_{10}{(t'}) + 2.87,
\label{eqn:D50}
\end{equation}
for circular, co-planar orbits.  The separation required for non-circular and/or mutually inclined orbits is given by
\begin{equation}
D_{50} \approx D_{50}(0,0) + \left(\frac{\left< e \right>}{0.01}\right) + \left(\frac{\left< i \right>}{0.04}\right),
\label{eqn:D50_00}
\end{equation}
where $D_{50}(0,0)$ is the value obtained at zero eccentricity and mutual inclination, defined by equation~(\ref{eqn:D50}); $\left< e \right>$ and $\left< i \right>$ are the typical values of eccentricity and inclination in the system.\\

Our {\sc{mercury6}} simulations exploring the stability of the GJ~887 system indicate that the 2 planet solution obtained with the REAL Gaussian Processes Kernel is stable across the posterior probability distribution of solutions. The 3 planet solution, however, is frequently unstable over run times of $10^5$ years. If we insert the parameters of the 3 planet Kepler solution 
into equations~(\ref{eqn:D50}) and (\ref{eqn:D50_00}), assume a coplanar system with $\left< i \right>=0$, and take the value $\left< e \right> = 0.18$ as the mean of the nominal values of the eccentricities for the three planets, then we obtain $D_{50}=25.48$. In other words, the mutual separations between neighbouring planets in the system ought to be $\sim 25 R_{\rm H}$ in order for the system to be stable for $10^5$ years. The nominal 3 planet solution has  $R_{\rm H} \sim 17$ for the inner planet pair, and $R_{\rm H} \sim 19$ for the outer pair, indicating that stability over simulation run times of $10^5$ is only expected for low eccentricity systems, in agreement with the {\sc{mercury6}} simulation outcomes. \\

\subsubsection*{Collisional evolution of unstable planetary systems}
The solutions obtained for the GJ~887 system from the RV data are consistent with the inner planet having a small eccentricity ($e_{\rm b} = 0.09^{+0.09}_{-0.06}$), and with GJ~887-c having a larger eccentricity $e_{\rm c}=0.22^{+0.09}_{-0.10}$. If planet d exists, then its eccentricity is $e_{\rm d} =0.25^{+0.20}_{-0.15}$ from the posterior probability distributions for the 3 planet solution. The mutual separations of $\sim 17 R_{\rm H}$ and $\sim 19 R_{\rm H}$ are consistent with earlier evolution that may have involved gravitational scattering and collisions among a larger number of planets. In a compact system such as GJ~887, where the planets are close to the central star and hence located deep within its gravitational potential, the evolution is unlikely to involve objects being scattered out of the system, but instead we expect it to involve collisions within a planetary system that becomes dynamically unstable.  Whether scattering or collisions dominate is determined by the Safranov number 
\begin{equation}
\Theta^2=\left(\frac{m_p}{M_*}\right)\left(\frac{a_{\rm p}}{R_{\rm p}}\right),
\label{eqn:Safranov}
\end{equation}
where $m_p$ is the mass of a planet, $R_p$ is the radius of a planet and $a_p$ is the semi-major axis. The Safranov number is related to the ratio of the escape velocity from the surface of a planet to its orbital velocity. Scattering is favoured in a system when $\Theta > 1$, whereas collisions are favoured when $\Theta< 1$. The planetary radii are unknown for GJ~887, so we assume a mean internal density $\rho=3$~g~cm$^{-3}$. With the parameter values for planets b, c, (and a putative d), Equation~(\ref{eqn:Safranov}) gives values in the range 0.17 -- 0.37, so collisions would be strongly favoured for such a compact system. \\

We can assess the likely outcome of this collisional evolution, and the expected range of orbital eccentricities. Gravitational scattering excites orbital eccentricities and inclinations, whereas inelastic collisions damp them. N-body simulations of in situ planetary accumulation for semi-major axes in the range $0.1 \le a_{\rm p} \le 1$ au indicate that planets can end up with final eccentricities $e \sim 0.2$ \cite{2013ApJ...775...53H}. The in situ formation of more compact systems, similar to GJ~887, suggests that 80\% of planets end up with $e \le 0.1$, and only 20\% have eccentricities in the range $0.1 \le e \le 0.2$ \cite{2020MNRAS.491.5595P}. An earlier phase of collisional evolution in the GJ~887 system would favour the lower eccentricity solutions arising in the posterior probability distributions, but higher eccentricity outcomes are not ruled out.\\

\subsubsection*{Tidal evolution}
\begin{figure}
\includegraphics[width=15cm]{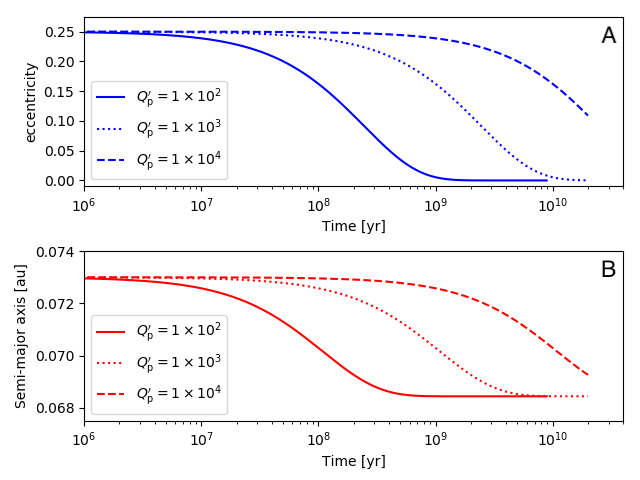}
\caption{
{\bf Tidal evolution of GJ~887-b's eccentricity and semi-major axis.} The top panel shows the eccentricity versus time, and the bottom panel shows the semi-major axis versus time, for the different values of $Q^{\prime}_{\rm p}$ indicated in the legends.
}
\label{fig:tidalevolution}
\end{figure}

The architecture of the GJ~887 planetary system, with orbital spacing in the range $\sim 17$ - $19 \, R_{\rm H}$, is consistent with a prior phase of dynamical instability. This would be expected to yield  moderately eccentric orbits. The eccentricity of GJ~887-c is consistent with this, but GJ~887-b probably has a small eccentricity $e_{\rm b} \le 0.09$. Since GJ~887\,b orbits close to the star, it may have experienced subsequent tidal circularisation. We quantified this process by integrating the tidal evolution equations for eccentricity and semimajor axis \cite{2004ApJ...610..464D}, assuming aligned stellar and planetary spins and conservation of orbital angular momentum. Estimates for the values of the tidal dissipation parameters, $Q_{\rm p}'$, for Solar System planets range between $100 \le Q_{\rm p}' \le 10^6$, with higher values applying to the gas giant planets and lower values applying to terrestrial bodies \cite{1966Icar....5..375G, 1981Icar...47....1Y, 2016CeMDA.126..145L}. We adopt a value of the stellar tidal dissipation parameter, $Q_{*}' \simeq 10^6$, derived from circularisation times in stellar clusters \cite{1998ApJ...502..788T}. We examined the tidal evolution for values of $Q_{\rm p}'$ in the range $100 \le Q_{\rm p}' \le 10^4$, i.e., values appropriate for rocky planets, super-Earths and Neptune-like bodies. The evolutionary tracks for the resulting eccentricities and semimajor axes are shown in Fig.~\ref{fig:tidalevolution} as a function of $Q_{\rm p}'$. We find that for $Q_{\rm p}' \le 10^3$ the planet evolves onto an essentially circular orbit, whereas for $Q_{\rm p}' = 10^4$ the tidal evolution is slow and GJ~887-b would remain on an eccentric orbit if it had been subjected to gravitational scattering earlier in the history of the system.

\begin{table}
\center
\caption{\label{tab:data_properties} {\bf{Radial velocity observations.}} Listed are the numbers of measurements ($N$), data baselines ($\Delta T_{\rm obs}$), standard deviations about the mean ($\sigma_{\rm SD}$), average instrument noises ($\langle\sigma\rangle$), and standard deviation of the residuals.  The last is not necessarily a measure for the instrument performance in an analysis using an inhomogeneous data set (see text for more details).  HARPS arc indicates HARPS archive observations, including data taken before and after the fiber upgrade.}
\begin{tabular}{lcccccccl}
\hline \hline
Data set & Year & Wavelength & $N_{\rm obs}$ & $\Delta T_{\rm obs}$ & $\sigma_{\rm SD}$ & $\langle\sigma\rangle$ & $\sigma_{\rm SD\, res}$ & Program\\
&  & range nm && d & ms$^{-1}$ & ms$^{-1}$ & ms$^{-1}$ & ID/Survey\\
\hline
HARPS new & 2018 & 378--691  & 65 & 82 & 3.48 & 0.1& 1.03 & 101.C-0516\\
&&&&&&&& 101.C-0494\\ 
&&&&&&&& 102.C-0525\\
HARPS arc & 2013-2017 & 378--691 & 72 & 4909 & 3.62 & 0.5& 1.66 & 072.C-0488\\
&&&&&&&& 096.C-0499\\
&&&&&&&& 098.C-0739\\
&&&&&&&& 099.C-0205\\
&&&&&&&& 100.C-0487\\
&&&&&&&& 191.C-0505\\
&&&&&&&& 192.C-0224\\
PFS & 2011-2013 & 391--734  & 38 & 827 & 4.83 & 2.2& 2.45 & Magellan \cite{Teske2016AJ....152..167T}\\
&&&&&&&& Planet\\
&&&&&&&& Search\\
HIRES &1998-2013 & 364--782 & 75 & 5655 & 4.83 & 0.9& 2.43 & HIRES/Keck\\
&&&&&&&& Exoplanet\\
&&&&&&&& Survey \cite{Butler_2017}\\
UCLES & 1998-2012 & 390--700 & 38 & 5106 & 4.59 & 1.4& 2.55 & Anglo-\\
&&&&&&&& Australian\\ 
&&&&&&&& survey \cite{Tinney2001ApJ...551..507T}\\
\hline
Combined & 1998-2018 & & 288 & 7406 & 3.67 & & 1.39 \\
\hline
\end{tabular}
\end{table}

\begin{table}
    \caption{\label{tab:phot-obslog} {\bf{Properties of the photometric data.}} Listed are the time span ($\Delta T_{\rm obs}$), number of individual observations ($N_{\rm obs}$), number of nights ($N_n$) and rms as average uncertainty over all nights in each data set. The latter is given for the nightly averaged data for ASH2.}
\vspace{0.4cm}
    \centering
    \begin{tabular}{lcrrrr}
        \hline
        \hline
        Data set     &  Year  & $\Delta T_{\rm obs}$ &$N_{\rm obs}$ &  $N_n$   &    rms~~~~~~\\
                     &           &    [d]     &           &          &      [mmag] \\
        \hline
        ASH2 B       &     2018  &   96.7      &  700      & 32      &  4          \\
        ASAS-3 V & 2002-2004 & 855.8 & 154 & 154 & 10\\
        TESS         &    2018 &  27.4      &   18317   &  --    &     0.3     \\
        \hline
    \end{tabular}
\end{table}

\begin{table}
\center
\caption{\label{tab:mag-activity} {\bf {Periodicities in stellar activity indicators and photometric data.} } The corresponding periodograms are shown in Fig.~\ref{fig:periodograms}.  Listed are the spectral ranges and pass-bands used.  For the S-index calculation the values for $H$ and $K$ are the normalised flux at the line cores, using a triangular pass-band, and $V$ and $R$ are the nearby continuum regions (respectively referred to as line 1, line 2, continuum region 1, continuum region 2). The lower panel gives the periodicities in the photometric data. S + NaD is the S-index + NaD.}
\begin{tabular}{lcccccccl}
\hline \hline
Index/ & line 1 & line 2 & pass-band   & continuum & continuum & P & Amp. \\
Photom. &  (nm)  &  (nm)  & width (nm) & No 1 (nm) & No 2 (nm) & (day) & ($\Delta \ln  L$) \\ 
\hline
S-index   & 393.363 & 396.847 & 1.09 & 389.1--391.1 & 399.1--401.1 & 55.8 & 8.75  \\
H$\beta$  & 486.136 & -- & 7.00 & 484.2--484.8           & 489.3--489.9   & 35.5 & 5.07  \\
Na\,D     & 588.995 & 589.592 & 3.75 & 584.0--585.0   & 592.5--593.5   & 54.9 & 12.43 \\ 
H$\alpha$ & 656.280 & -- & 7.00 & 644.2--644.8           & 657.6--658.0 & 37.9 & 9.41  \\
H$\alpha$ + H$\beta$ &-- &--&--&--&--& 37.0 & 14.1 \\
S + NaD  &-- &--&--&--&--& 54.9 & 15.0 \\
\hline
ASH2 & -- & -- & 110.0 (B) & -- & -- & -- & -- \\
ASAS & -- & -- & 99.1 (V) & -- & -- & 194.7 &  12.8 \\
TESS & -- & -- & 400 & -- & -- & 13.7$^{*}$ & 16.1\\
\hline
\end{tabular}
\tablefoot{$^{*}$ Caution is advised in interpreting this very low amplitude periodic signal as the stellar rotation period.}
\end{table}

\begin{table}
    \caption{{\bf{Detection and model comparison table}}. The signals are listed in order of detection using likelihood
    periodograms. The period of the signals included in the model are given for reference. (*) When using the REAL kernel to
    model correlated noise, the solution has an almost identical likelihood as without the 3rd Keplerian and the period of the
    third signal becomes poorly constrained. Note that in all cases, the models using the REAL kernel substantially improve those
    without Gaussian processes (GP).}
    \vspace{0.4cm}
    \centering
    \begin{tabular}{lllll}
\hline
\hline
Parameter               & nosignals  &  1 Keplerian & 2 Keplerians & 3 Keplerians\\            \hline
$P_\mathrm{1}$ [d]      & --         & 21.8         & 21.8         &  21.8       \\
$P_\mathrm{2}$ [d]      & --         &  --          &  9.2         &   9.2       \\
$P_\mathrm{3}$ [d]      & --         &  --          &  --          &  50.7       \\
\hline
$\ln L_{\rm no GP}$        & -847	     & -814	        & -760	       & -729  \\
$\delta \ln L_{\rm no GP}$ & 0	         &  +43	        & +54	       &  +31  \\
\hline
$\ln L_{\rm REAL}$     & -782	     & -769	        & -698	       & -698(*) \\
$\delta \ln L_{\rm REAL}$
                       & 0	         & +13	        &  +71	       &    0(*) \\
\hline
$\ln L_{\rm REAL} - \ln L_{\rm no GP}$ 
                       & +65	     & +45	        &  +62	       &    +31 \\
\hline
\hline
    \end{tabular}
    \label{tab:modelcomparison}
\end{table}

\begin{table}[]
    \centering
    \caption{{\bf Priors for the model parameters of the best-fit model.} }
    \begin{tabular}{lccr}
       \noalign{\smallskip}
       \hline
        \hline
        \noalign{\smallskip}
         Parameter & Prior & Units & Description \\
         \hline
         P$_b$ & $\mathcal{U}(9.2,9.3)$ & d & orbital period\\
         P$_c$ & $\mathcal{U}(21.7,21.9)$ & d & orbital period\\
         K$_{b,c}$ & $\mathcal{U}(0,100)$ & m\,s$^{-1}$ &RV semi amplitude \\
         e$_{b,c}$ & $\mathcal{U}(0,1)$ & & eccentricity of orbit\\
         $\omega_{b,c}$ & $\mathcal{U}(-\infty,\infty)$ & rad & argument of periastron\\
         t$_{0,b,c}$ & $\mathcal{U}(-\infty,\infty)$ & d & time of periastron\\
         Offsets & $\mathcal{U}(-\infty,\infty)$ & m\,s$^{-1}$ &instrumental offsets\\
         Jitter & $\mathcal{LU}(-15,10)$ & m\,s$^{-1}$ &instrumental jitter values\\
         a & $\mathcal{LU}(-10,4)$ & m$^2$\,s$^{-2}$ &variance of REAL kernel \\
         c & $\mathcal{LU}(-5,5)$ & d$^{-1}$&inverse life time of REAL Kernel \\
\hline
    \end{tabular}
    \label{tab:priors}
\end{table}

\begin{table}
    \centering
    \caption{{\bf{Jitter and Offsets}}. The resulting jitter and offset terms for all instruments. For HARPS, HIRES and UCLES, the posterior distribution of the jitter parameter  is a one-sided distribution, we therefore list the 95\% percentile value} 
    \vspace{0.4cm}
    \begin{tabular}{lll}
        \hline
        \hline
         Instrument & Jitter & Offset \\
         \hline
         HARPS pre [m s$^{-1}$]  & $<1.0 $        &$1.4\pm 1.2$\\
         HARPS post [m s$^{-1}$] & $<0.6 $        &$0.5\pm 1.2$\\
         PFS [m s$^{-1}$]        & $2.4\pm 0.7$ &$0.7\pm 1.2$\\
         HIRES [m s$^{-1}$]      & $<1.8 $        &$2.4\pm 1.2$\\
         UCLES [m s$^{-1}$]      & $<3.1 $        &$3.2\pm 1.4$\\
         \hline
    \end{tabular}
    \label{tab:jitteroffset}
\end{table}

\begin{table}
    \caption{\label{tab:activity_corr} {\bf{Correlations with the stellar activity indicies}}. Listed are the Pearson's r-coefficients and the student's t-test stp-values. }
    \centering
    \vspace{0.4cm}
    \begin{tabular}{lcc}
        \hline
        \hline
Pairs of activity indicies & Pearsons (r) & Student's t-test (stp) \\
        \hline
H$\alpha$ vs H$\beta$ &  0.89  & 2.06$\times$10$^{-21}$ \\
S-index vs NaD & 0.93 & 1.15$\times$10$^{-27}$ \\ 
H$\alpha$ vs S-index & 0.57 & 2.10$\times$10$^{-6}$\\
H$\alpha$ vs NaD & 0.56 & 3.59$\times$10$^{-6}$\\
H$\beta$ vs S-index & 0.71 & 2.00$\times$10$^{-10}$\\
H$\beta$ vs NaD &  0.73 &  3.10$\times$10$^{-11}$ \\
    \hline
RV vs H$\alpha$ & -0.11 &  0.45 \\
RV vs H$\beta$ &  -0.13 & 0.38 \\
RV vs S-index & 0.24 & 9.07$\times$10$^{-2}$ \\
RV vs NaD & -0.24 & 0.10\\
\hline
    \end{tabular}

\end{table}

\newpage

\end{document}